\documentclass[11pt,preprintnumbers]{article}
\usepackage{jheppub}
\usepackage{bbm}
\usepackage{epsfig,bm}
\usepackage{graphics,graphicx,comment,subfigure}
\usepackage{amsmath}
\usepackage{amsthm}
\usepackage{color}
\usepackage{mathtools}
\usepackage{multirow} 
\usepackage{bigstrut}

\usepackage[toc,nonumberlist]{glossaries}
 
\makeglossaries

 \newglossaryentry{Formal}
{
    name={$\mathbb{C}[[z^{-1}]]$},
    description={Set of formal power series in $1/z$},
    sort= a
}

 \newglossaryentry{Germ}
{
    name={$\mathbb{C}\{\zeta\}$},
    description={Set of convergent power series in $\zeta$, i.e. germs of analytic functions at the origin},
        sort= b
}

\newglossaryentry{phitilde}
{
    name={$\tilde{\phi}(z)$},
    description={Generic physical observable as a formal power series in $1/z$ },
        sort= bb
}
 
\newglossaryentry{Borel}
{
    name={$\mathcal{B}[\tilde{\phi}]$},
    description={Borel transform of the formal power series $\tilde{\phi}$},
        sort= bo
}

\newglossaryentry{phihat}
{
    name={$\hat{\phi}(\zeta)$},
    description={Borel transform of a generic formal power series $\tilde{\phi}$ },
        sort= bob
}
 
\newglossaryentry{laplace}
{
    name={$\mathcal{L}^\theta\hat{\phi}(\zeta)$},
    description={Directional Laplace transform of $\hat{\phi}$ along the complex direction $\arg = \theta$} ,
           sort= c
}

\newglossaryentry{Htilde}
{
    name={$\tilde{\mathcal{H}}$},
    description={Multiplicative model of the algebra of resurgent functions},
        sort= f
}

\newglossaryentry{Hhat}
{
    name={$\hat{\mathcal{H}}$},
    description={Convolutive model of the algebra of resurgent functions},
        sort= g
}

\newglossaryentry{Restilde}
{
    name={$\widetilde{\mbox{RES}}^{simp}$},
    description={Multiplicative model of the algebra of simple resurgent functions},
        sort= h
}

\newglossaryentry{Reshat}
{
    name={$\widehat{\mbox{RES}}^{simp}$},
    description={Convolutive model of the algebra of simple resurgent functions},
        sort= i
}

\newglossaryentry{Sing}
{
    name={$\mbox{Sing}_\omega \hat{\phi}(\zeta)$},
    description={Singular part of the simple resurgent function $\hat{\phi}(\zeta)$ close to the point $\omega$},
        sort= il
}

\newglossaryentry{S}
{
    name={$\mathcal{S}_{\theta^\pm} \tilde{\phi}$},
    description={Lateral Borel sum of the formal power series $\tilde{\phi}$ along the complex direction $\arg = \theta$},
        sort= j
}

\newglossaryentry{Stokes}
{
    name={$\mathfrak{S}_{\theta} \tilde{\phi}$},
    description={Stokes automorphism of the formal power series $\tilde{\phi}$ along the complex direction $\arg = \theta$},
        sort= k
}

\newglossaryentry{Alien}
{
    name={$\Delta_\omega \tilde{\phi} $},
    description={Alien derivative of the formal power series $\tilde{\phi}$ at the singular point $\omega$},
        sort= l
}

\newglossaryentry{Disc}
{
    name={$\mbox{Disc}_\theta \Phi(z)$},
    description={Discontinuity of the analytic function $\Phi(z)$ across the complex direction $\arg =\theta$},
        sort= m
}

\DeclarePairedDelimiter{\floor}{\lfloor}{\rfloor}

\newcommand{\pslash}{\not{\hbox{\kern-2.3pt $p$}}}
\newcommand{\qslash}{\not{\hbox{\kern-2.3pt $q$}}}
\newcommand{\kslash}{\not{\hbox{\kern-2.3pt $k$}}}
\newcommand{\partialslash}{\not{\hbox{\kern-2.3pt $\partial$}}}
 
\theoremstyle{remark}
\newtheorem{Def}{Definition}

\theoremstyle{remark}
\newtheorem{Ex}{Example}

\theoremstyle{definition}
\newtheorem{Prop}{Proposition}

\theoremstyle{definition}

\def \be { \begin{equation} }
\def \ee { \end{equation} }

\def\Re{{\rm Re}\,}

\def\Dslash{{\rlap{\raise 1pt \hbox{$\>/$}}D}}

\def \( {\left(}
\def \) {\right)}

\def \bDef{\begin{definition}}
\def \eDef{ \end{definition} }

\preprint{{\flushright DAMTP-2014-44\\}}

\title{An Introduction to Resurgence, Trans-Series and Alien Calculus}

\author{Daniele Dorigoni}
\emailAdd{d.dorigoni@damtp.cam.ac.uk}
\affiliation{%
DAMTP, University  of Cambridge, 
Wilberforce Road, Cambridge CB3 0WA, UK
}

\abstract{
In these notes we give an overview of different topics in resurgence theory from a physics point of view, but with particular mathematical flavour.
After a short review of the standard Borel method for the resummation of asymptotic series, we introduce the class of simple resurgent functions, explaining their importance in physical problems. We define the Stokes automorphism and the alien derivative and discuss these objects in concrete examples using the notion of trans-series expansion.
With all the tools introduced, we see how resurgence and alien calculus allow us to extract non-perturbative physics from perturbation theory. To conclude, we apply Morse theory to a toy model path integral to understand why physical observables should be resurgent functions.}

\begin{document}

\maketitle
\clearpage
\glsaddall
\printglossary[title=List of Mathamatical Symbols and Notations,toctitle=Glossary]
\clearpage

\section{Introduction}
\label{sec:Intro}

When confronted with computing various physical quantities, i.e. partition functions, vacuum energies, anomalous dimensions of operators, in different physical systems, we are almost always facing the problem that, unless something miraculous is coming to help (integrability, supersymmetric localization,...)
we will not be able to get an exact answer.
So either we sit idle and declare defeat or we try one of the few things that we are almost always\footnote{The $6$ dimensional $\mathcal{N}=(2,0)$ is precisely an exception to this.} able to do: perturbation theory.

To perform a perturbative expansion, we first have to find a suitable parameter, say $g$, that we can tune to be small, and as we dial it from zero to some specific value, we interpolate between a simpler model, at $g=0$, for which we have an exact answer for the physical observable $\mathcal{O}$ under consideration, and the actual model of interest.
At this point we expect to be able to write our observable as a power series in $g$
\begin{equation}
\mathcal{O}(g)=c_0+ c_1\,g+c_2\,g^2+...\,,\label{eq:perturbIntro}
\end{equation}
where $c_0$ is the same observable we are interested in but computed in the exactly solvable model at $g=0$ and all the correction $c_n$ can be in principle computed within this exact model.

We are all familiar with the astonishing precision test of QED perturbation theory used to compute the anomalous $g-2$ magnetic moment of the electron and, in the quantum mechanics course, we all computed the corrections to the energy levels of the harmonic oscillator in the presence of a quartic perturbation. What is probably less appreciated is that if we were to compute higher and higher corrections we would encounter bigger and bigger contributions.

In QFT, a standard argument by Dyson \cite{Dyson:1952tj} suggests that the perturbative expansion is only an asymptotic expansion and must have vanishing radius of convergence for $g\sim 0$. The origin of the asymptotic character of perturbation theory is the rapid growth of Feynman diagrams \cite{Hurst:1952zh,Bender:1976ni}: the number of diagrams contributing at order $n$ grows factorially with $n$. This combinatorial argument by itself is not enough to conclude that the perturbative series is asymptotic, some magic cancelations might happen when we sum all the diagrams contributing to a certain order. Thanks to Lipatov \cite{Lipatov:1976ny} we know that this is not the case: one can show, via a saddle-point method, that indeed the perturbative coefficients grow at higher orders as $c_n\sim n!\,$. Similarly in quantum mechanics \cite{Bender:1969si,ZinnJustin:1980uk,Bender:1990pd}, matrix models \cite{Marino:2007te,Pasquetti:2009jg} and topological strings \cite{Marino:2006hs}, we encounter the same factorial growth of the coefficients $c_n$ in (\ref{eq:perturbIntro}), effectively making our perturbative expansion only an asymptotic series \cite{Dingle_asymptotics} with zero radius of convergence.

It was soon realised that the asymptotic nature of the perturbative expansion was actually hiding deep and valuable information about the exact answer.
The venerable idea of the Borel summation was introduced as a suitable analytic continuation of our asymptotic series by means of a contour integral of the associated Borel transform in the complex plane.
This procedure generically gives rise to ambiguities in the resummation process due to the presence of poles in the Borel transform, changing the contour of integration leads to many different analytic continuations of the \textit{same} physical observables $\mathcal{O}(g)$ (i.e. Stokes phenomena).
The ``strength" of these ambiguities is related to terms that cannot be possibly captured by an expansion of the form (\ref{eq:perturbIntro}), precisely the non-perturbative (NP) physics.

Furthermore, even in cases when the Borel sum of the perturbative series alone would give rise to an unambiguous analytic continuation, this might not be the exact answer \cite{Grassi:2014cla}. We have to investigate the analytic properties of the Borel transform in the entire complex Borel plane. We stress that, in general, we do not have a complete argument\footnote{See that last Section for a possible explanation.} for why the poles of the Borel transform of the perturbative expansion should all be associated with new NP physics so it is perhaps surprising that, in all the cases analysed in the literature, it is always possible to find a suitable weak coupling regime in which these poles can be interpreted as particular non-perturbative objects of the underlying microscopic theory, i.e. instantons, D-branes, quasi-normal modes \cite{Heller:2013fn} etc.

It is clear then that if we want a unique and well defined resummation procedure for our observable $\mathcal{O}(g)$, we have to use something more general than (\ref{eq:perturbIntro}).
A natural extension of the perturbative expansion is the so called trans-series expansion\footnote{We present it here in its most basic form, see later on for a more general and complete discussion.}
\begin{equation}
\mathcal{O}(g) = \sum_{n\geq 0} c^{(0)}_n g^n + \sum_i e^{-S_i/g} \sum_{n\geq 0} c_n^{(i)} g^n\,.\label{eq:transIntro}
\end{equation}
We note that the terms $e^{-S_i/g}$ are precisely the type of terms that cannot be captured by a perturbative expansion for $g$ small, since they vanish, together with all their derivatives, for $g\to 0$.
From a path integral point of view we are just adding to the perturbative expansion around the vacuum, $\sum c_n^{(0)} g^n$, the (multi-)instantons corrections $e^{-S_i/g}$, or other type of non-trivial saddle points, and the perturbative expansions, $\sum c_n^{(i)} g^n$, on top of them.
As for the original perturbative series, all these new perturbative expansions, around all the different non-perturbative saddles, will only be asymptotic series. If we Borel transform each one of the series $\sum c_n^{(i)} g^n$ and try to apply our resummation procedure, as described before, we will introduce many more new ambiguities: naively it looks like we only made our original problem worse.

Luckily for us there is a systematic mathematical framework, called resurgence theory, to study precisely these kind of trans-series.
Resurgence theory was discovered in a different context by Ecalle in the early 80s \cite{Ecalle:1981} and since then it has been applied with success to quantum mechanics \cite{ZinnJustin:1980uk,Voros:1983,Voros:1993an,Delabaere:1999ef,Jentschura:2004jg}, matrix models \cite{Marino:2012zq}, supersymmetric localizable field theories \cite{Aniceto:2014hoa} and topological string theory \cite{Aniceto:2011nu,Santamaria:2013rua,Couso-Santamaria:2014iia}.
Only recently Argyres, Dunne and \"Unsal \cite{Argyres:2012vv,Argyres:2012ka,Dunne:2012ae,Dunne:2012zk} were able to apply resurgence to certain asymptotically free QFT and they were able to obtain, for the first time, a weak coupling interpretation of the IR renormalons \cite{'tHooft:1977am,Beneke:1998ui}.

Resurgent functions can be describe using a certain class of trans-series with particular distinctive properties.
The key properties, as we will see in more details later on, are that each perturbative expansion $\sum c_n^{(i)} g^n$, appearing in the trans-series, will define, through its Borel transform, a holomorphic function with ``few" singularities in the complex Borel plane and whose behaviour close to its critical points will be entirely captured by the Borel transform of the perturbative expansion associated to a different perturbative series $\sum c_{n}^{(j)} g^n$, i.e. the coefficients $c_n^{(i)}$ know about different non-perturbative saddles.
To put it in a suggestive way the perturbative series around the trivial vacuum will know of all the other non-perturbative saddle points $e^{-S_i/g}$ and it will also know about their perturbative series coefficients $c_n^{(i)}$ and vice versa.

The idea behind these notes is to give a short introduction to resurgence, by keeping to a minimum the number of mathematical technicalities and always having in mind concrete examples. For a more rigorous and mathematical discussion we refer to the comprehensive books by Ecalle \cite{Ecalle:1981} and the more recent and excellent works by Sauzin \cite{Sauzin1,Sauzin2}.

In Section \ref{sec:Borel} we will give an overview of the standard Borel transform and resummation procedure, leading, in Section \ref{sec:Resurge}, to the introduction of the algebra of simple resurgent functions.
The behaviours of this class of resurgent functions close to their singular points is the subject of Section \ref{sec:Stokes},
where we will introduce the Stokes automorphism and the alien derivative.
In Section \ref{sec:Trans} we will briefly discuss some generic properties of trans-series. With all the tools introduced we will be able to understand, in Section \ref{sec:Bridge}, how the perturbative and non-perturbative physics are linked together by means of the bridge equations. 
Finally, in Section \ref{sec:Median}, we will combine everything together and discuss a path integral toy model: Borel Ecalle theory will tell us how to perform an unambiguous median resummation and, in Section \ref{sec:Out}, we will conclude with some speculation on why physical observables should take the form of simple resurgent functions.

\section{Borel Resummation}
\label{sec:Borel}

With a slight change of notation from the physics literature and the Introduction, instead of describing a physical observable as a formal series obtained from a weak coupling expansion $g \sim  0$, we will write everything in terms of $z=1/g$ and work at $z\sim \infty$.
 By denoting with $\mathbb{C}[[z^{-1}]]$, the set of all the formal power series (generically with infinitely many terms) in $1/z$ ,
we can introduce the algebra of formal power series for $z\sim \infty$
\begin{equation}
z^{-1} \mathbb{C}[[z^{-1}]] = \left\lbrace \sum_{n=0}^\infty c_n z^{-n-1} \,,\,c_n\in \mathbb{C}\right\rbrace\,.
\end{equation} 
Every formal power series is specified by an infinite list $\{c_n\}$ of complex numbers that can be seen as the coefficients of a infinite order polynomial in $1/z$ without constant term\footnote{The reason to avoid the constant term is just a technicality and we will remove this restriction later on.}.

\begin{Def}
We can define a linear operator $\mathcal{B}$ called \textit{Borel Transform}
\begin{align}
&\mathcal{B} : \,z^{-1} \mathbb{C}[[z^{-1}]] \rightarrow \mathbb{C}[[\zeta]]\,,\\
&\label{eq:convB}\mathcal{B}: \tilde{\phi}(z)= \sum_{n=0}^\infty c_n z^{-n-1}\rightarrow \hat{\phi}(\zeta) =\sum_{n=0}^\infty c_n\frac{\zeta^n}{n!}\,.
\end{align}
\end{Def}
This operator improves the convergence of the original series, in fact if $\tilde{\phi}(z)$ converges for all $\vert z^{-1} \vert < r$, its Borel transform $\hat{\phi}$ is an entire function of exponential type in every direction
\begin{equation}
\vert \hat{\phi}(\zeta) \vert \leq C e^{ R \vert \zeta \vert }
\end{equation}
with $R> r$. Conversely if $\hat{\phi}=\mathcal{B}[\tilde{\phi}]$ has only a finite radius of convergence the radius of convergence of $\tilde{\phi}$ will be zero.
It is easy to see how basic properties of the formal power series $\tilde{\phi}\in z^{-1}\mathbb{C}[[z^{-1}]]$ are translated into properties of its Borel transform
\begin{align}
&\mathcal{B}:\,\partial \tilde{\phi}(z) \rightarrow - \zeta \hat{\phi}(\zeta)\,,\\
&\mathcal{B}:\,\tilde{\phi}(z+1)\rightarrow e^{-\zeta} \hat{\phi}(\zeta)\,,\\
&\mathcal{B}:\,\tilde{\psi}(z)\,\tilde{\phi}(z)\rightarrow (\hat{\psi}\ast\hat{\phi})(\zeta)=\int_0^\zeta d\zeta_1\,\hat{\psi}(\zeta_1)\,\hat{\phi}(\zeta-\zeta_1)\,.
\end{align}
The last property is telling us that when we pass to the Borel transform, the natural multiplication of formal power series in the algebra $z^{-1}\mathbb{C}[[z^{-1}]]$ becomes a convolution in $\mathbb{C}[[\zeta]]$. This is why sometimes, when we will compute physical observables as asymptotic power series, we will say that they belong to the formal \textit{multiplicative model}, while when we will pass to their Borel transforms we will be working in the \textit{convolutive model}.
The precise mapping between different type of operations, in the multiplicative model and in the convolutive model, is concisely  presented in Table \ref{tab}\footnote{We thank Mithat \"Unsal for this Table.}.

\begin{table}[t] 
\begin{center}
\begin{tabular}{|c|c|}
\hline
Multiplicative model & Convolutive model  \bigstrut \\
\hline
$ \tilde{\phi}(z) $  &$  \hat{\phi}(\zeta)$ \bigstrut \\
\hline
 $ z^{-\alpha-1} $  &$  \zeta^{\alpha}/\,\Gamma (\alpha+1)$  \bigstrut\\
\hline
$ \partial_z \tilde{\phi}(z)$ &$ - \zeta\, \hat{\phi}(\zeta)$  \bigstrut  \\
\hline
$z \,\tilde{\phi}(z)$ & $ \partial_\zeta
\hat{\phi}(\zeta)$  \bigstrut \\
\hline
$ \tilde{\phi}(\lambda z) $  &$  \lambda\, \hat{ \phi} (\lambda^{-1}
\zeta) $ \bigstrut \\
\hline
$\tilde{\phi}(z+\lambda)$ & $e^{-\lambda \zeta} \hat{\phi}(\zeta)$  \bigstrut \\
 \hline
$ \tilde{\psi}(z)\,\tilde{\phi}(z) $ & $
(\hat{\psi}\ast\hat{\phi})(\zeta)=\int_0^\zeta
d\zeta_1\,\hat{\psi}(\zeta_1)\,\hat{\phi}(\zeta-\zeta_1)$ \bigstrut \\
\hline
\end{tabular}
\end{center}
\caption{Mapping of operations from the multiplicative model to the convolutive model. }
\label{tab}
\end{table}

Just to clarify better: if we have two observables $\tilde{\phi}_1(z),\tilde{\phi}_2(z)$ obtained as formal asymptotic power series, their product will be once again a formal asymptotic power series
\begin{equation}
\tilde{\phi}_1(z)\,\tilde{\phi}_2(z) = \left( \sum_{n=0}^\infty a_n z^{-n-1} \right) \left( \sum_{n=0}^\infty b_n z^{-n-1} \right) =  \left( \sum_{n=1}^\infty c_n z^{-n-1} \right)\,,
\end{equation}
where the coefficients $c_n$ are simply the convolution sum of the $\{a_n\}$ and $\{b_n\}$
\begin{equation}
c_n = \sum_{p+q=n-1}a_p\, b_q\,,\qquad n\geq 1\,.\label{eq:cn}
\end{equation}

Passing to the Borel transforms $\hat{\phi}_1(\zeta),\hat{\phi}_2(\zeta)$ we can write their convolution product (\ref{eq:convB}) as
\begin{align}
(\hat{\psi}\ast\hat{\phi})(\zeta)&\notag=\sum_{n,m\geq0} \frac{a_n \,b_m}{n!\,m!}\int_0^\zeta d\zeta_1\,\zeta_1^n (\zeta-\zeta_1)^m\\
&=\sum_{n,m\geq0} \frac{a_n \,b_m}{n!\,m!} B(n+1,m+1)\,\zeta^{n+m+1}\,,
\end{align}
where $B(n,m)$ is the Euler Beta function.
We can rearrange the convolution product by using the known identity $B(n,m)=\Gamma(n)\Gamma(m)/\Gamma(n+m)$, obtaining
 \begin{equation}
(\hat{\psi}\ast\hat{\phi})(\zeta) = \sum_{n,m\geq 0} \frac{a_n \,b_m}{(n+m+1)!} \zeta^{n+m+1}=\sum_{n\geq 1} \frac{c_n}{n!} \zeta^n\,,
\end{equation}
with precisely the same coefficients $c_n$ found before in (\ref{eq:cn}) from the product of the two formal series $\tilde{\phi}_1,\tilde{\phi}_2$.
Hence the name multiplicative model for the algebra of formal power series and convolutive model for their Borel transforms.

Standard observables in QFT, when computed in a perturbative regime, takes precisely the form $\tilde{\phi}(z) = \sum_{n=0}^\infty c_n z^{-n-1}$, with $1/z $ playing the role of the small coupling constant, with the tree-level contribution subtracted out (see later on in this Section). As already mentioned in the Introduction, thanks to standard arguments \cite{Dyson:1952tj,Bender:1976ni,Lipatov:1976ny}, we know that, since the number of Feynman diagrams at order $n$ grows factorially with $n$, the coefficients $c_n = O(C^n\, n!)$ will diverge and the perturbative expansion will only be an asymptotic expansion, with zero radius of convergence.
\begin{Def}
We will say that a formal power series $\tilde{\phi}(z)=\sum_{n=0}^\infty c_n z^{-n-1}$ is of \textit{Gevrey order-}$1/m$, if the large orders asymptotic terms are bounded by
\begin{equation}
\vert c_n \vert \leq \alpha\,C^n\,( n!)^m\,,
\end{equation}
for some constants $\alpha$ and $C$.
\end{Def} 
Note that thanks to Stirling formula the Gevrey order of a formal power series with $c_n \sim (n!)^m$ is the same of the power series with $d_n\sim (m\cdot n)!$.
From the arguments of Dyson and Lipatov, we can deduce that, in standard QFT, physical observables computed by perturbation theory are given by Gevrey-1 formal power series.
\begin{Prop}
The Borel transform $\hat{\phi}(\zeta)$ of a formal power series $\tilde{\phi}(z)=\sum_{n=0}^\infty c_n z^{-n-1}$ has a finite radius of convergence if and only if $\tilde{\phi}(z)$ is of Gevrey-1 type
\begin{equation}
\vert c_n \vert = O( C^n\,n!)\,.
\end{equation}
\end{Prop}
Note that if the $c_n$ are growing faster than $n!$, for example $c_n \sim (2n)!$, we can always make a change of variables $z\to z(w)$, in this case $z\to w^2$, so that the new series in $w^{-n-1}$ is of Gevrey-1 type, however this change of variables will introduce new monodromies in the complex $z$-plane. 

From now on we will always assume, unless differently specified, that $\tilde{\phi}$ is of Gevrey-$1$\footnote{
Note that, generically, the coefficients $c_n$, obtained from the perturbative expansions of our favourite physical quantity, contain as well power law and logarithmic corrections to the leading factorial growth.
Roughly speaking they take the form $c_n\sim n! \,n^\alpha\,\log^\beta n$ which is clearly of Gevrey-$1$ type.} type and its Borel transform $\hat{\phi}$ defines a convergent expansion at the origin, in mathematical language it defines a germ of analytic functions at $\zeta \sim 0$. A germ of analytic functions at $z_0$ is the set of all the analytic functions with the same Taylor expansion around the point $z_0$.
We will usually write a germ $\hat{\phi}$ of analytic functions at the origin as $\hat{\phi}\in\mathbb{C}\{ \zeta \}$.

After having improved the convergence of the original formal series $\tilde{\phi}\to \mathcal{B}[\tilde{\phi}]$, we need an operator to bring us back to a suitable analytic extension of the original formal series. 
\begin{Def}
We define the \textit{directional Laplace transform}:
\begin{equation}
\mathcal{L}^\theta [\hat{\phi}](z) = \int_0^{e^{i \theta} \infty} d\zeta\,e^{-z \,\zeta} \hat{\phi}(\zeta)\,.
\end{equation}
\end{Def}
This operator is linear and it maps analytic functions on $e^{i\theta}\mathbb{R}^+ $, with rate of growth of at most exponential type $e^{r \vert \zeta \vert}$, into analytic functions $\mathcal{L}^\theta \hat{\phi}$ in the half plane $\Re \left(z e^{i \theta}\right)> r$.

In particular, let's note that we can easily compute $\mathcal{L}$ on the real positive line for all the monomials 
\begin{equation}
\mathcal{L}^0\,[\zeta^\alpha]= \frac{\Gamma(\alpha+1)}{z^{\alpha+1}}\,,\label{eq:Lapzeta}
\end{equation}
and similarly for the inverse Laplace transform
\begin{equation}
\left(\mathcal{L}^0\right)^{-1} \,[z^{-\alpha-1}]=\frac{\zeta^\alpha}{\Gamma(\alpha+1)}\,.\label{eq:Lapz}
\end{equation}
Thank to Table \ref{tab}, we see that $\mathcal{L}^0$, when applied to $\zeta^\alpha$, acts precisely as the inverse Borel transform.

\begin{Ex}
Euler studied the properties of the series 
\begin{equation}
\tilde{\phi}(z)=\sum_{n=0}^{\infty} (-1)^n n!\,z^{-n-1}\label{eq:EulerAlt}
\end{equation}
for $z=1$. He noticed that $\tilde{\phi}(z)$ formally solves the ODE
\begin{equation}
\phi'(z)-\phi(z)=-\frac{1}{z}\,.\label{eq:Euler}
\end{equation}
From (\ref{eq:EulerAlt}) it is easy to compute the Borel transform of $\tilde{\phi}$ obtaining
\begin{equation}
\hat{\phi}(\zeta) =\frac{1}{1+\zeta}\,.
\end{equation}
By applying the Laplace transform in the direction $\theta=0$, we get the analytic continuation in the half-plane $\Re(z)>0$ of the formal series $\tilde{\phi}(z)$
\begin{equation}
\mathcal{L}^0[\hat{\phi}] (z) = \int_0^\infty d\zeta\,e^{-z\,\zeta}\frac{1}{1+\zeta}= e^{z} \,\Gamma(0;\,z)\,,
\end{equation}
where $\Gamma(0;z)$ is the incomplete gamma function.
If we expand the above equation for $z \to \infty$ we recover the formal power series defined above, furthermore it is easy to check that $\mathcal{L}^0\hat{\phi} (z)$ is a particular solution to Euler's equation (\ref{eq:Euler}), while the generic solution takes the form
\begin{equation}
\varphi(z)=e^{z} \,\Gamma(0;\,z)+ C\,e^z\,,
\end{equation}
with $C$ an arbitrary constant. The homogeneous term $C\,e^z$ is non-analytic for $z\to \infty$ and cannot be expanded as a formal power series in $z^{-1}\,\mathbb{C}[[z^{-1}]]$, these kind of terms will be explored more in details in the context of trans-series, see Section \ref{sec:Trans}.
In this case, the Borel sum of the asymptotic power series $\tilde{\phi}(z)$ gives us the unique solution, $\varphi(z)$, to Euler's equation (\ref{eq:Euler}), vanishing at $z\to\infty$.
\end{Ex}

\begin{figure}
\begin{center}
	\includegraphics[scale=0.3]{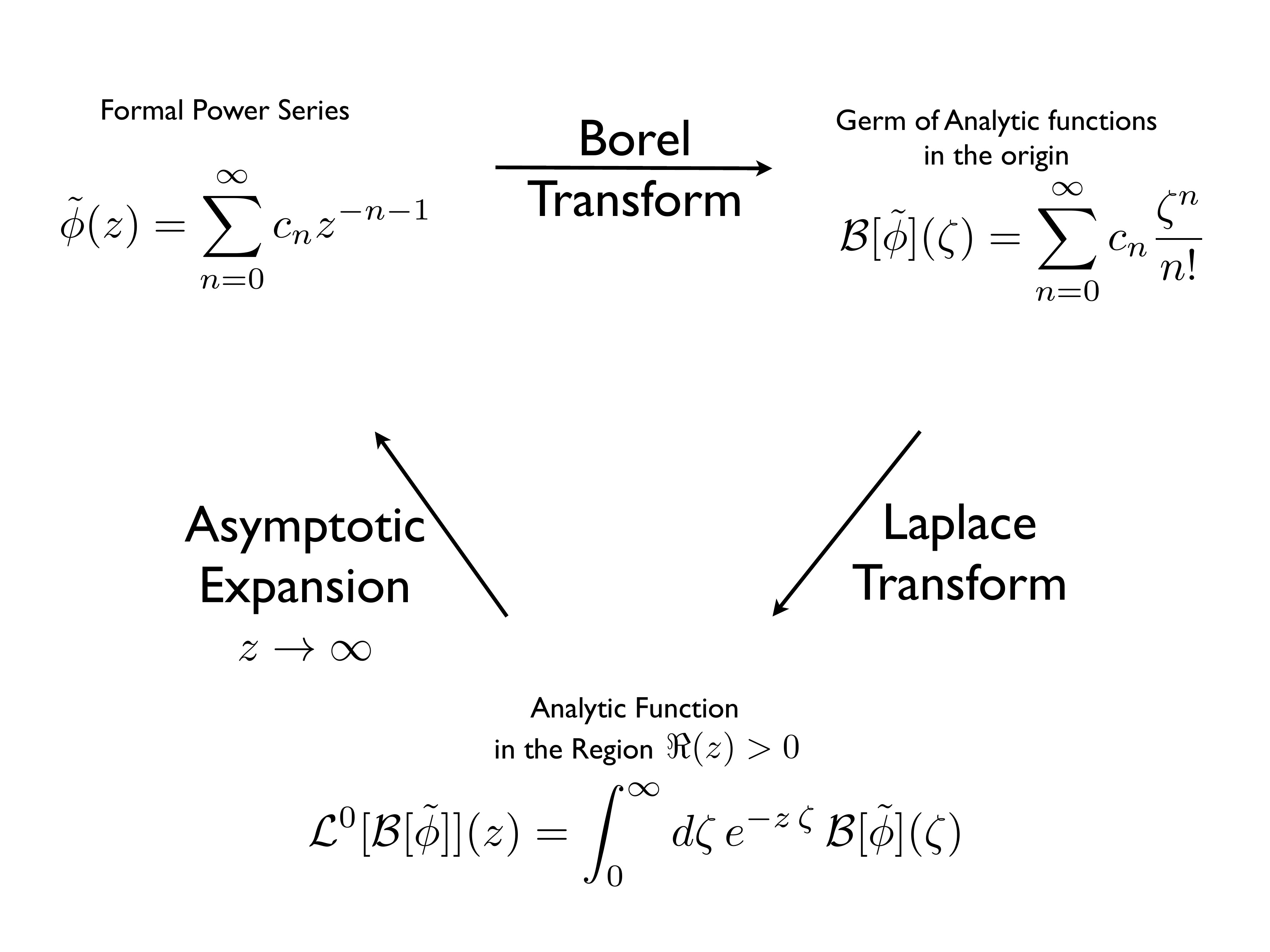}
	\caption{Schematic form of the Borel regularisation procedure.}
	\label{Fig:scheme}
\end{center}
\end{figure}
The idea behind combining Borel transform and Laplace transform comes from the very well known equation for the gamma function
\begin{equation}
1=\frac{1}{n!} \int_0^\infty d\zeta\,\zeta^{n}\,e^{-\zeta} \,,
\end{equation}
by plugging this identity, or using (\ref{eq:Lapzeta})-(\ref{eq:Lapz}), in each term of the formal power series $\tilde{\phi}$ we get (after a trivial change of variables)
\begin{equation}
\tilde{\phi}(z) =\sum_{n=0}^\infty  \int_0^\infty d\zeta\, e^{-z\,\zeta}\, \frac{c_n}{n!} \, \zeta^{n}\,.
\end{equation}
We can commute the sum with the integral and obtain an analytic continuation of our original diverging series as the Laplace integral of its Borel transform
\begin{equation}
\tilde{\phi}(z) = \mathcal{L}^0 [ \mathcal{B}[\tilde{\phi}]](z)\,,
\end{equation} 
this defines for us a regularisation procedure for our diverging series, see Figure \ref{Fig:scheme}.

In the example above we were able to compute exactly the Laplace integral for the Borel transform of the formal power series $\tilde{\phi}$, solution to Euler's equation, but generically, unless $\hat{\phi}$ is analytic along the contour of integration, we will not be able to do so. The resurgent functions are a particular class of formal series for which the singularities in the $\zeta$-plane (also called \textit{Borel plane}) will satisfy certain conditions. Just by studying the behaviour of such functions close to their singular points, we will be able to constrain, via the Alien calculus discussed in Section \ref{sec:Bridge}, the entire structure of the function on the whole Borel plane.

\section{The Algebra of Simple Resurgent Functions}
\label{sec:Resurge}

As we have already mentioned, only in very few cases the Borel transform $\hat{\phi}=\mathcal{B}[\tilde{\phi}]$ will not have any singularity along the line of integration, and even in rarer occasions there will be no singularity at all (and actually in this situation there is no need for all this machinery). So, generically, we will be expecting at least some singularity in $\hat{\phi}$. The number and type of such singularities is encoded in the following definitions. First of all we will need to be able to integrate along some path from the origin to infinity, so we cannot have ``too many" singularities.
\begin{Def}
We will say that a germ of analytic functions at the origin $\hat{\phi}\in\mathbb{C}\{\zeta\}$ is \textit{endlessly continuable} on $\mathbb{C}$ if for all $R > 0$, there exists a finite set $\Gamma_R(\hat{\phi})\subset\mathbb{C}$ of accessible singularities, 
such that $\hat{\phi}$ can be analytically continued along all paths $\gamma$ whose length is less than $R$, avoiding the singularities $\Gamma_R(\hat{\phi})$.
\end{Def}
Ecalle's definition is more general than the one just presented, but for the present work this definition will suffice.
Being endlessly continuable means that even if the Borel transform of our formal power series will present possibly infinitely many singularities in the Borel plane, nonetheless it will be possible to consider a suitable deformed path $\gamma$, issuing from the origin and going to infinity in any direction $\theta$. Endless continuability roughly means that there are no natural boundaries in the, possibly infinitely sheeted, Riemann surface where $\hat{\phi}$ is defined.

We have to assume as well some hypothesis on the type of singularities that $\hat{\phi}$ can have.
\begin{Def}
A holomorphic function $\hat{\phi}$ in an open disk $D\subset \mathbb{C}$ is said to have a \textit{simple singularity} at $\omega$, adherent\footnote{I.e. $\omega$ belongs to the closure of $D$.} to $D$, if there exist $\alpha \in \mathbb{C}$ and two germs of analytic functions at the origin $\hat{\Phi}(\zeta),reg(\zeta)\,\in\mathbb{C}\{\zeta\}$, such that
\begin{equation}
\hat{\phi}(\zeta) = \frac{\alpha}{2\pi i \,(\zeta-\omega)}+\frac{1}{2\pi i}\hat{\Phi}(\zeta-\omega)\log (\zeta-\omega)+ reg(\zeta-\omega)\,,\label{eq:SimpleSing}
\end{equation}
for all $\zeta \in D$ close enough to $\omega$, where $reg$ stands for a regular term close to $\omega$.
The constant $\alpha$ is called the \textit{residuum} and $\hat{\Phi}$ the \textit{minor}.
\end{Def}
The holomorphic function $\hat{\Phi}$ associated with the logarithmic singularity can be obtained by considering
\begin{equation}
\hat{\Phi}(\zeta) = \hat{\phi}(\zeta+\omega) -\hat{\phi}(\zeta \,e^{-2 \pi i}+\omega)\,,
\end{equation}
where with $\hat{\phi}(\zeta e^{-2 \pi i}+\omega)$ we mean following the analytic continuation of $\hat{\phi}$ along the circular path $\zeta e^{-2\pi i\,t}+\omega$ with $t\in[0,1]$.

Endlessly continuable functions with simple singularities are stable under the natural operation of convolution, however this does not mean that the convolution of two such functions preserve the location and type of singularities. Convolution of functions with simple singularities generically generates multi-valuedness and new singularities.
\begin{Ex}
Consider in example the following convolution product
\begin{equation}
1 \ast \hat{\phi}(\zeta) = \int_0^\zeta d \zeta_1 \, \hat{\phi}(\zeta_1)\,,
\end{equation}
with $\hat{\phi}$ a meromorphic function with poles in $\Gamma\subset \mathbb{C}^\ast$.
Then clearly $1\ast \hat{\phi}$ admits an analytic continuation along any path issuing from the origin whose support is not intersecting $\Gamma$.
This means that $1\ast \hat{\phi}$ is actually an holomorphic function defined on the universal covering of $\mathbb{C} \setminus \Gamma$ with only logarithmic singularities located precisely at the poles of $\hat{\phi}$.
\end{Ex}
\begin{Ex}
\label{ex:Conv}
It is easy to see that the convolution of two functions with simple singularities generates new singular points.
We can take
\begin{align}
\hat{\phi}(\zeta) &= \frac{1}{\zeta-\omega_1}\,,\\
\hat{\psi}(\zeta) &= \frac{1}{\zeta-\omega_2}\,,\\
\hat{\phi}\ast \hat{\psi}(\zeta) &= \frac{1}{\zeta- (\omega_1+\omega_2)} \left(\int_0^\zeta d\zeta_1\,\frac{1}{\zeta_1-\omega_1}+\int_0^\zeta d\zeta_1\,\frac{1}{\zeta_1-\omega_2}\right)\,.
\end{align}
The product $\hat{\phi}\ast \hat{\psi}$ has logarithmic singularities at $\omega_{1},\omega_2$ and a pole at $\omega_1+\omega_2$ (note that this pole is not on the first sheet), thus we can extend $\hat{\phi}\ast \hat{\psi}$ to a meromorphic function on the universal covering of $\mathbb{C}\setminus \{\omega_1,\omega_2\}$ with a simple pole at $\omega=\omega_1+\omega_2$ whose residue depends on the particular sheet considered.
\end{Ex}

As it will become clearer later in our discussion, for each problem that we wish to solve through resurgence, we have to find an infinite discrete subset of points $\Gamma\subset\mathbb{C}$ (usually a lattice), corresponding to the singular points of all the germs of analytic functions in play for our particular problem (i.e. $\Gamma = 2\pi i\,\mathbb{Z}\setminus \{0\}$ for certain difference equations , see \cite{Sauzin1,Sauzin2}, while for various $2$d QFTs \cite{Dunne:2012ae,Cherman:2013yfa} $\Gamma= S_0 \,\mathbb{Z}$, with $S_0 >0$).

Given this discrite subset of singular points, $\Gamma \subset \mathbb{C}$, there is a natural Riemann surface associated with the {\it{universal covering}} of $\mathbb{C}\setminus\Gamma$.
\begin{Def}
The Riemann surface $\mathcal{R}$ is the set of homotopy classes of paths with fixed extremities, starting from the origin and whose support is contained in $\mathbb{C}\setminus \Gamma$.
The \textit{covering map} $\pi$ is a mapping from $\mathcal{R}$ back to $\mathbb{C}\setminus \Gamma$ given by
\begin{align}
&\pi\,:\,\mathcal{R}\rightarrow \mathbb{C}\setminus \Gamma\,,\\
&\pi[c] = \gamma(1) \in \mathbb{C}\setminus \Gamma\,,
\end{align}
where $\gamma(t)$ is a particular representative of the equivalence class $c\in \mathcal{R}$, and $\gamma(1)$ correspond to its end point.
By pulling back with $\pi$ the complex structure of $\mathbb{C}\setminus \Gamma$ we get $\mathcal{R}$ as a Riemann surface.
The origin of $\mathcal{R}$ is the unique point corresponding to $\pi^{-1}[0]$, which corresponds to the homotopy class of the constant path.
\end{Def}

Without dwelling on too many technical details we can introduce holomorphic functions on $\mathcal{R}$ by simply taking a germ of holomorphic functions at the origin which admits an holomorphic continuation along any path whose support avoids $\Gamma$.
Since we are dealing only with endlessly continuable functions, this $\Gamma$ cannot be too ``dense".
The key point is that the convolution of germs induces a commutative and associative law on the space of holomorphic functions of $\mathcal{R}$ (see the beautiful works of Sauzin for more details \cite{Sauzin1,Sauzin2}). 
\begin{Def}
The space of all holomorphic functions on $\mathcal{R}$ endowed with the convolution product is an algebra called \textit{convolutive model of the algebra of resurgent functions} and usually denoted by $\widehat{\mathcal{H}}(\mathcal{R})$.
Considering the inverse Borel transform of these functions we get 
\begin{equation}
\tilde{\mathcal{H}} = \mathcal{B}^{-1} \left( \widehat{\mathcal{H}}(\mathcal{R})\right)
\end{equation} which is usually called \textit{multiplicative model of the algebra of resurgent functions}.
\end{Def}

There is no unity for the convolution product $\ast$ within $\widehat{\mathcal{H}}(\mathcal{R})$, for this reason we have to introduce a new symbol $\delta$ and extend the algebra in the following straightforward manner
\begin{align}
& \tilde{\phi}(z) = C+\sum_{n=0}^\infty c_n z^{-n-1}\,\in\,\mathbb{C}[[z^{-1}]]\,,\\
& \hat{\phi}(\zeta) = \mathcal{B}[\tilde{\phi}](\zeta) = C \,\delta+\sum_{n=0}^\infty \frac{c_n}{n!}\, \zeta^n\,\in \,\delta\,\mathbb{C} \oplus \mathbb{C}[[\zeta]]\,.
\end{align}
The convolution product is extended from $\mathbb{C}[[\zeta]]$ to $\delta\,\mathbb{C} \oplus \mathbb{C}[[\zeta]]$ by simply treating $\delta$ as a unity $\delta \ast \hat{\phi} =\hat{\phi}$.

It is possible to show that this algebra behaves nicely under composition as well.
More details and proofs for all these statements can be found in the original works by Ecalle \cite{Ecalle:1981} (see also the more recent \cite{Sauzin1,Sauzin2}). 

Within the whole algebra $\widehat{\mathcal{H}}(\mathcal{R})$ we can focus on resurgent functions $\hat{\phi}$ with a simple singularity at $\omega$, which means
\begin{equation}
\hat{\phi}(\zeta) = \frac{\alpha}{2\pi i \,(\zeta-\omega)}+\frac{1}{2\pi i}\hat{\Phi}(\zeta-\omega)\log (\zeta-\omega)+ reg(\zeta-\omega)\,,
\end{equation}
where $\hat{\Phi}$ and $reg$ are convergent series close to the origin.
We can thus define the operator
\begin{equation}
\mbox{Sing}_\omega \hat{\phi} = \alpha\,\delta+\hat{\Phi}\,\in\,\delta\,\mathbb{C} \oplus \mathbb{C}\{\zeta\}\,.
\label{eq:Sing}
\end{equation}
Note that a change in the determination of the logarithm gives rise only to a change in the regular part, $reg$, and not on $\alpha$ or $\hat{\Phi}$.
\begin{Def}
A \textit{simple resurgent function} $\hat{\psi}$ is such that $\hat{\psi} = c\,\delta+\hat{\phi}(\zeta)\in\,\delta\,\mathbb{C}\oplus\widehat{\mathcal{H}}(\mathcal{R})$  and for all $\omega\in\Gamma$, i.e. all the accessible singularities, and all the paths $\gamma(t)$, originating from $0$, avoiding $\Gamma$ and whose extremity lies in a disk $D$ close enough to $\omega$ (where close enough means that $\omega$ is the only singularity contained in $D$, i.e. $\{\omega\} = \Gamma \cap D$), the determination $\mbox{cont}_\gamma \hat{\psi}$ has a simple singularity at $\omega$.
Where $\mbox{cont}_\gamma \hat{\psi}$ is the determination of $\hat{\psi}$ obtained by analytic continuation of $\hat{\psi}$ along $\gamma$. This continuation is clearly analytic at least in any open disk containing the extremity of the path $\gamma$ and avoiding all the singular points in $\Gamma$ (remember that endlessly continuable functions cannot have too dense singular points).
\end{Def}

This following proposition summarises all the concepts introduced so far in this Section.
\begin{Prop}
The subspace of all simple resurgent functions, which usually is denoted by $\widehat{\mbox{RES}}^{simp}$, is a subalgebra of the convolution algebra $\delta\,\mathbb{C}\oplus\widehat{\mathcal{H}}(\mathcal{R})$.
\end{Prop}
In the multiplicative model the conjugate through Borel transform of $\widehat{\mbox{RES}}^{simp}$ will be denoted by
$\widetilde{\mbox{RES}}^{simp}$.
The main points to remember about simple resurgent functions are briefly summarised in what follows:
\begin{itemize}
\item The singular points are not too dense. We want to be able to integrate from $0$ to infinity along some complex direction by suitably dodging few singular points;
\item The singular behaviour close to these singular points is captured by yet another simple resurgent function;
\item These functions behave nicely under composition, convolution and Borel transform \cite{Sauzin1,Sauzin2}, i.e. they form a sub-algebra of the convolutive model.
\end{itemize}

\section{Stokes Automorphism and Alien Derivative}
\label{sec:Stokes}

\begin{figure}
\begin{center}
	\includegraphics[scale=0.8]{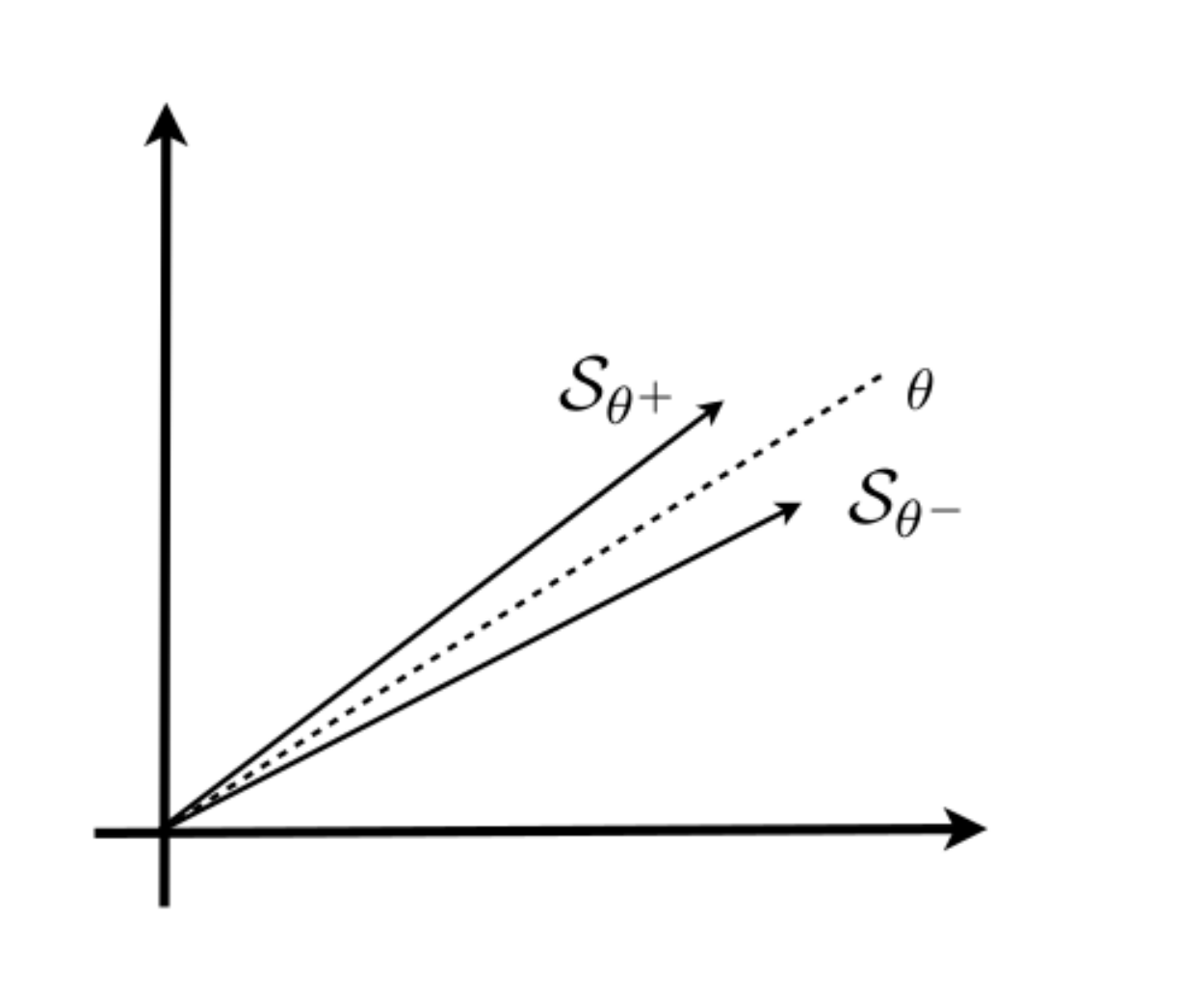}
	\caption{Lateral Borel summation along the direction $\theta$.}
	\label{Fig:Lateral}
\end{center}
\end{figure}

In the previous Section we introduced many formal concepts and properties of resurgent functions, but we still do not know how to define a suitable resummation procedure when the direction $\theta$, along which we compute the Laplace integral $\mathcal{L}^\theta$, contains singular points.
Since we cannot directly integrate along a singular direction, we have to ``dodge" the singular points.
\begin{Def}
The \textit{lateral Borel summations} for $\tilde{\psi} = c+\tilde{\phi}(\zeta)\in\mathbb{C}\oplus\tilde{\mathcal{H}}$ along the direction $\theta$ are given by
\begin{align}
\mathcal{S}_{\theta^+} \tilde{\psi}(z)&= c+\int_0^{e^{ i\,\theta}(\infty+i\, \epsilon)} d\zeta\,e^{-z\,\zeta} \,\hat{\phi}(\zeta)\,,\\
\mathcal{S}_{\theta^-} \tilde{\psi}(z)&= c+\int_0^{e^{ i\,\theta}(\infty-i\,\epsilon)} d\zeta\,e^{-z\,\zeta} \,\hat{\phi}(\zeta)\,,
\end{align}
as schematically depicted in Fig.\ref{Fig:Lateral}:
we deform slightly the contour of integration to pass either above ($ \mathcal{S}_{\theta^+} $) or belove ($\mathcal{S}_{\theta^-} $) all singular points along the direction $\theta$.
\end{Def}

\begin{Ex}
A slight modification to Euler's equation
\begin{equation}
\psi'(z)+\psi(z)=\frac{1}{z}\,,\label{eq:Euler2}
\end{equation}
yields to the following formal power series solution  $\tilde{\phi}(z)=\sum_{n=0}^{\infty} n!\,z^{-n-1}$.
Note the absence of the alternating factor $(-1)^n$ present instead in the formal solution to eq. (\ref{eq:Euler}).
The Borel transform is straightforward to obtain
\begin{equation}
\hat{\phi}(\zeta) =\frac{1}{1-\zeta}\,.
\end{equation}
We cannot apply directly the Laplace transform along the direction $\theta=0$ since we would encounter a singularity at $\zeta =1$ (the direction $\theta =\pi$ for Euler's equation (\ref{eq:Euler}) is exactly conjugated to the direction $\theta=0$ for the current ODE).
The later Borel summations differ from each others
\begin{align}
&\mathcal{S}_{+} \tilde{\phi}(z) = \int_0^{\infty+i\,\epsilon} d\zeta\,\frac{e^{-z\,\zeta}}{1-\zeta}\,,\\
&\mathcal{S}_{-} \tilde{\phi}(z) = \int_0^{ \infty-i\,\epsilon} d\zeta\,\frac{e^{-z\,\zeta}}{1-\zeta}\,,
\end{align}
and
\begin{equation}
(\mathcal{S}_{+} -\mathcal{S}_{-}) \tilde{\phi}(z) = 2\pi i\,e^{-z}\,.
\end{equation}
Note that the difference between the two lateral summations is non-analytic for $z\sim \infty$ and cannot be possibly captured by our formal asymptotic expansion in powers of $1/z$ (see Section \ref{sec:Trans}). Secondly the difference $e^{-z}$ is precisely a solution to the homogenous problem $\psi'(z)+\psi(z)=0$.
\end{Ex}

When the direction $\theta$ is a singular direction, the Borel summation jumps as we cross this \textit{Stokes line},
and the full discontinuity across this direction plays a crucial role.
\begin{Def}
Consider the lateral Borel summations $\mathcal{S}_{\theta^\pm}$, we define the \textit{Stokes automorphism} $\mathfrak{S}_\theta$, from $\widetilde{\mbox{RES}}^{simp}$ into itself, as
\begin{align}
&\mathcal{S}_{\theta^+} = \mathcal{S}_{\theta^-} \circ \mathfrak{S}_\theta = \mathcal{S}_{\theta^-}\circ \left(\mbox{Id}-\mbox{Disc}_\theta\right)\label{eq:Stokes}\,,\\
&\mathcal{S}_{\theta^+}-\mathcal{S}_{\theta^-}=-\mathcal{S}_{\theta^-}\circ \mbox{Disc}_\theta\,.
\end{align}
Where $\mbox{Disc}_\theta$ encodes the full discontinuity across $\theta$.
\end{Def}

If $\mathcal{S}_{\theta^+}(\tilde{\phi})=\mathcal{S}_{\theta^-}(\tilde{\phi})$ we easily obtain
\begin{equation}
\mathfrak{S}_\theta \,\tilde{\phi} = \tilde{\phi}\,,
\end{equation}
and $\tilde{\phi}$ is called a \textit{resurgence constant}. This means that the Borel transform of $\tilde{\phi}$ has no singularities along the $\theta$ direction and is given by a convergent power series. In this case we have already seen that the Laplace integral of the Borel transform of $\tilde{\phi}$ gives us an unambiguous resummation procedure for the original formal power series.

The Stokes automorphism is telling us how the resummed series jumps across a Stokes line, as we will see in more details later the reason for this jump is that our formal power series $\tilde{\phi}(z)=c_0+c_1/z+c_2/z^2+...$ is actually incomplete, we missed non-analytic (non-perturbative) terms of the form $e^{-z}$. These terms are of course exponentially suppressed for $z\sim \infty$ but across a Stokes line, precisely the terms that we have forgotten, become relevant and have to be taken into account.

\begin{figure}
\begin{center}
	\includegraphics[scale=0.45]{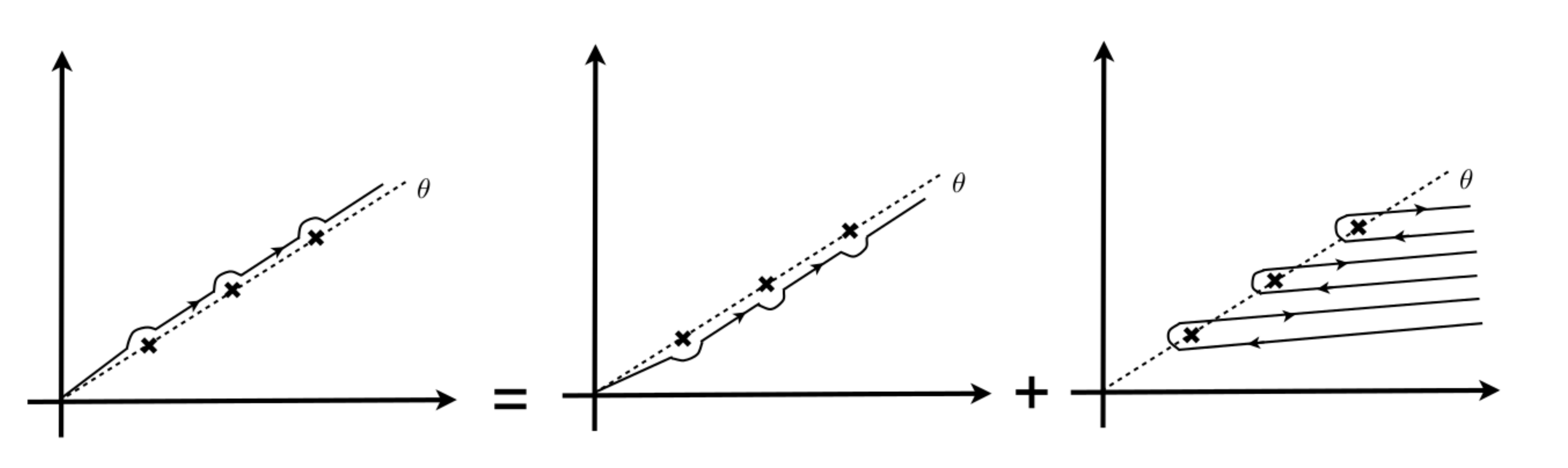}
	\caption{The difference between left and right resummation along the singular direction $\theta$ as a sum over Hankel contours.}
	\label{Fig:Stokes}
\end{center}
\end{figure}

It is easy to see, by a simply contour deformation, that the difference between the $\theta^+$ and $\theta^-$ deformation is nothing but a sum over Hankel's contours, and the discontinuity of $\mathcal{S}$ across $\theta$ is given as an infinite sum of contribution coming from each one of the singular points, see Figure \ref{Fig:Stokes}.

\begin{Def}
The logarithm of the Stokes automorphism defines the \textit{Alien derivative} $\Delta_\omega$ by
\begin{equation}
\mathfrak{S}_\theta = \exp \left(\sum_{\omega\in\Gamma_\theta} e^{-\omega\,z}\Delta_{\omega}\right)\,,\label{eq:expAlien}
\end{equation} 
where we denoted with $\Gamma_\theta$ the set of singular points of the Borel transform along the $\theta$ direction.
\end{Def}
Using the above definition we can rewrite equation (\ref{eq:Stokes}) as
\begin{equation}
\mathcal{S}_{\theta^+} \tilde{\phi}(z) = \mathcal{S}_{\theta^-} \tilde{\phi}(z)+\sum_{k=1}^\infty \sum_{\{n_1,...n_k\geq 1\}}\frac{e^{-(\omega_{n_1}+...+\omega_{n_k})\,z}}{k!} \,\mathcal{S}_{\theta^-} \left(\Delta_{\omega_{n_1}}...\,\Delta_{\omega_{n_k}}\,\tilde{\phi}(z)\right)\,.\label{eq:SAlien}
\end{equation}

The Alien, \textit{etranger}, derivative can be thought of as the logarithm of the Stokes automorphism, but our definition (\ref{eq:expAlien}) is still pretty mysterious and unintelligible.

\begin{Ex}
To understand better how this Alien derivative works we can start with the easier task of understanding the Stokes automorphism when the Borel transform of our formal power series $\tilde{\phi}(z)\in\widetilde{\mbox{RES}}^{simp}$ takes the form
\begin{equation}
\hat{\phi}(\zeta)=\frac{\alpha}{2\pi i\,(\zeta-\omega)}+\frac{1}{2\pi i} \hat{\Phi}(\zeta-\omega)\,\log (\zeta-\omega)\,,
\label{eq:1Sing}
\end{equation}
with $\arg(\omega)=\theta$ and $\hat{\phi}$ has no singularities along the direction $\theta$.
Note that we are assuming the simple singularity form (\ref{eq:SimpleSing}) along the whole direction $\theta$, and not only close to the singular point $\omega$.
The difference between the two lateral resummations is clearly different from zero and it gets a first contribution coming from the simple pole and a second one coming from the change in the determination of the logarithm.
Having assumed that $\hat{\Phi}$ is entire along the direction $\theta$, after a trivial change of variables $\zeta\to\zeta-\omega$ ,we get
\begin{equation}
(\mathcal{S}_{\theta^+}-\mathcal{S}_{\theta^-}) \tilde{\phi}(z) = \alpha\,e^{-\omega\,z} + e^{-\omega\,z}\,\int_0^\infty d\zeta\,e^{-z\,\zeta}\,\hat{\Phi}(\zeta)\,.
\end{equation}
Since $\omega$ is the only singular point, equation (\ref{eq:SAlien}) simplifies drastically to
\begin{equation}
(\mathcal{S}_{\theta^+}-\mathcal{S}_{\theta^-}) \tilde{\phi}(z) = e^{-\omega\,z} \,\mathcal{S}_{\theta^-} \left(\Delta_\omega\, \tilde{\phi}(z)\right)\,,
\end{equation}
from which we can read how the alien derivative act on a simple resurgent function of the form (\ref{eq:1Sing})
\begin{equation}
\Delta_\omega\, \tilde{\phi}(z) = \alpha+ \tilde{\Phi}(z)\,,\label{eq:AlienSimple}
\end{equation}
where $\tilde{\Phi}$ is the inverse Borel transform of $\hat{\Phi}$, or equivalently in the convolutive model
\begin{equation}
\Delta_\omega\, \hat{\phi}(\zeta) = \alpha\,\delta+ \hat{\Phi}(\zeta)\,.
\end{equation}
\end{Ex}

\begin{Ex}
Let's consider a more concrete example. Take the formal power series
\begin{equation}
\tilde{\phi}(z) = \frac{a}{z\,\omega}+\frac{a+b-c}{(z\,\omega)^2} + \sum_{n=2}^\infty \frac{n!}{(z\,\omega)^{n+1}}\left(a+\frac{b}{n}+\frac{c}{n(n-1)}\right)\,,
\end{equation}
with $a,b,c,\omega\in\mathbb{C}$ external parameters. This series is clearly of Gevrey type 1 and it is not so difficult to compute its Borel transform
\begin{equation}
\hat{\phi}(\zeta) = \frac{-a}{\zeta-\omega}- \frac{c (\zeta-\omega)+b\, \omega}{\omega^2} \,\log (1-\zeta/\omega)\,.
\end{equation}
The residuum at $\omega$ is $\alpha = -2 \pi i\,a$ while the minor $\hat{\Phi}(\zeta) = -2\pi i (c \,\zeta+b \omega)/\omega^2$.
In this case it is particularly easy to find the inverse Borel transform of the minor $\tilde{\Phi}(z) = -2\pi i(b/ (z\,\omega)+c/ (z\,\omega)^2)$, so that the Alien derivative at $\omega$ is 
\begin{equation}
\Delta_\omega \,\tilde{\phi}(z) = -2\pi i\left(a+ \frac{b}{(\omega \,z)}+\frac{c}{(\omega \,z)^2}\right)\,.\label{eq:AlienEasy}
\end{equation}
\end{Ex}

\begin{figure}
\begin{center}
	\includegraphics[scale=0.8]{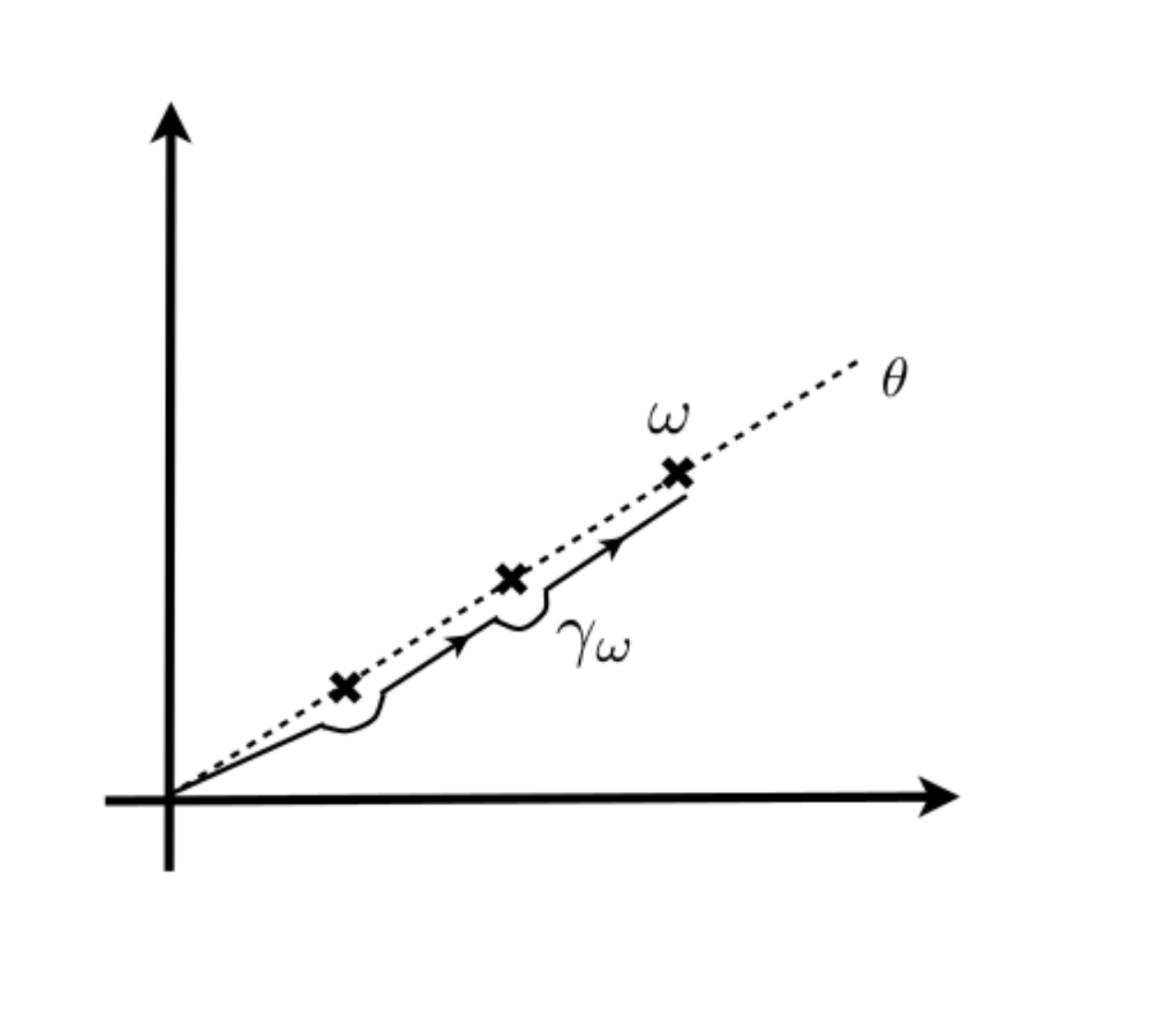}
	\caption{To obtain $\Delta_\omega^+$, we have to consider the determination of $\hat{\phi}$ along the path $\gamma_\omega$, issuing from the origin and reaching $\omega$ by avoiding all the singularities from the right.}
	\label{Fig:AlienRight}
\end{center}
\end{figure}

In the generic case the definition of the Alien derivative is a little bit more complicated, but the main idea are collected in the previous example. Along a singular direction the alien derivative at a singular point will receive contributions from all the singularities it encounters along its way
\begin{equation}
\Delta_\omega = \sum_n\,\sum_{\omega_1+...+\omega_n = \omega}\frac{(-1)^{n-1}}{n} \Delta_{\omega_1}^+\,...\,\Delta_{\omega_n}^+\,,\label{eq:AlienPlus}
\end{equation}
where $\omega_i$ are singular points, ordered along the direction $\theta$ and the operator $\Delta_{\omega}^+$ from $\widehat{\mbox{RES}}^{simp}$ into itself, is defined by
\begin{equation}
\Delta_\omega^+ \hat{\phi}(\zeta) = \alpha_{\gamma_\omega} \,\delta+\hat{\Phi}_{\gamma_{\omega}}(\zeta)\,,
\end{equation}
where $\alpha_{\gamma_\omega}\in\mathbb{C}$ and $\hat{\Phi}_{\gamma_{\omega}}\in\widehat{\mbox{RES}}^{simp}$ are respectively the residuum and the minor of $\hat{\phi}$ at the simple singularity $\omega$, as in equation (\ref{eq:SimpleSing}), defined following the determination of $\hat{\phi}$ along the path $\gamma_\omega$, issuing from the origin in the direction $\theta=\arg (\omega)$ and arriving in $\omega$ by circumventing all the intermediate singularities to the right, see Figure \ref{Fig:AlienRight}. For an equivalent definition without having to introduce $\Delta_\omega^+$ see \cite{Sauzin1,Sauzin2}.

Thanks to the Borel transform, we can keep on moving from the convolutive model to the multiplicative formal model, so we can understand $\Delta_\omega$ as acting on both $\widehat{\mbox{RES}}^{simp}$ and $\widetilde{\mbox{RES}}^{simp}$.
As the name suggest the Alien derivative is indeed a derivative
\begin{align}
&\Delta_\omega \left(\hat{\phi}_1\ast\hat{\phi}_2\right)=\Delta_\omega\,\hat{\phi}_1 \ast \hat{\phi}_2+\hat{\phi}_1\ast\Delta_\omega\, \hat{\phi}_2\,,\\\label{eq:Jacobi}
&\Delta_\omega \left(\tilde{\phi}_1\cdot \tilde{\phi}_2\right)=\Delta_\omega\, \tilde{\phi}_1 \cdot \tilde{\phi}_2+\tilde{\phi}_1\cdot\Delta_\omega\, \tilde{\phi}_2\,.
\end{align}
Note that however, $\Delta_\omega^+$ is generically not a derivation.
The alien derivative does not commute with the standard derivative but rather
\begin{equation}
\Delta_\omega \, \partial_z\,\tilde{\phi} = \partial_z\,\Delta_\omega \tilde{\phi}-\omega\,\Delta_\omega \tilde{\phi}\,.
\end{equation}
We can define the dotted alien derivative
\begin{equation}
\dot{\Delta}_\omega = e^{-\omega\,z}\Delta_\omega\,,
\end{equation}
which, thanks to the previous equation, does now commute with the standard derivative
\begin{equation}
\left[ \partial_z,\dot{\Delta}_\omega\right]=0\,.
\end{equation}
Note that $e^{-\omega\,z}$ has to be understood as a new symbol (see Section \ref{sec:Trans}), external to the algebra of simple resurgent functions,
this is usually called a \textit{simple resurgent symbol}. It obeys the usual rules for multiplication and derivation with respect to $z$. These simple resurgent symbols can be used to obtain the graded algebra $\widetilde{\mbox{RES}}^{simp}[[e^{-\omega\,z}]]$, where $\omega \in \Gamma$ are all the singular points for the particular problem studied.
The introduction of these symbols is telling us that somehow our formal power series expansion has to be extended to a more general expansion, which goes under the names of trans-series expansion, we refer to Section \ref{sec:Trans} for more details.
For the time being we just need to know that $e^{-\omega\,z}$ has to be understood as an external symbol to our algebra of simple resurgent functions, hence
\begin{equation}
\dot{\Delta}_{\omega_1} \, \left(e^{-\omega_2\,z}\,\tilde{\phi}\right) = e^{-(\omega_1+\omega_2)\,z}\Delta_{\omega_1} \,\tilde{\phi}\,.
\end{equation}

\textbf{Remark.} We have to stress that there is no operatorial relations between the various $\Delta_{\omega}$: they generate a free Lie algebra. They give a way to encode the entire singular behaviour of a resurgent function $\hat{\phi}$: in fact, given a sequence $\omega_1,...,\omega_N$ of singular points, the evaluation of $\Delta_{\omega_1}\,...\,\Delta_{\omega_N}\, \hat{\phi}$ is obtained by many different determinations of $\hat{\phi}$ at the singularity $\omega=\omega_1+...+\omega_N$. Vice versa any possible determination and singularity of $\hat{\phi}$ could be computed if we knew all these compositions of alien derivatives for all the sequences $\omega_1,...\omega_N$.

\begin{Ex}
The full knowledge of a resurgent function is coming only when we know ALL its alien derivatives. In particular it is not sufficient to know that $\Delta_\omega \,\hat{\phi}=0$ to deduce that $\tilde{\phi}$ has no singularities at $\omega$ for all its determinations. In fact, let's analyse the previous Example \ref{ex:Conv}
\begin{align}
\hat{\phi}(\zeta) &\notag= \frac{1}{\zeta-\omega_1}\,,\qquad\hat{\psi}(\zeta) = \frac{1}{\zeta-\omega_2}\,,\\
\hat{\phi}\ast \hat{\psi}(\zeta) &\notag= \frac{1}{\zeta- (\omega_1+\omega_2)} \left(\int_0^\zeta d\zeta_1\,\frac{1}{\zeta_1-\omega_1}+\int_0^\zeta d\zeta_1\,\frac{1}{\zeta_1-\omega_2}\right)\,.
\end{align}
We can easily compute the alien derivatives
\begin{align}
&\Delta_{\omega_1} \hat{\phi} = 2\pi i\,\delta\,,\qquad\Delta_{\omega} \hat{\phi} =0\, \qquad\forall\, \omega \neq \omega_1\,,\\
&\Delta_{\omega_2} \hat{\psi} = 2\pi i\,\delta\,,\qquad\Delta_{\omega} \hat{\psi} = 0\,\qquad\forall\, \omega \neq \omega_2\,,
\end{align}
so that
\begin{equation}
\Delta_{\omega_1+\omega_2} (\hat{\phi}\ast \hat{\psi}) =(\Delta_{\omega_1+\omega_2} \hat{\phi})\ast \hat{\psi}+\hat{\phi}\ast (\Delta_{\omega_1+\omega_2} \hat{\psi})=0  \,,
\end{equation}
where we used the Jacobi identity (\ref{eq:Jacobi}). The vanishing of $\Delta_{\omega_1+\omega_2} (\hat{\phi}\ast \hat{\psi})$ does not mean that all the determinations of $\hat{\phi}\ast \hat{\psi}$ have no singularity at $\omega_1+\omega_2$ as it is manifest from the explicit form for $\hat{\phi}\ast \hat{\psi}$. 
This fact is encoded in the composition of different alien derivatives
\begin{align}
&\Delta_{\omega_1} (\hat{\phi}\ast \hat{\psi})= 2\pi i\,\hat{\psi}\,,\qquad \Delta_{\omega_2} (\hat{\phi}\ast \hat{\psi})= 2\pi i\,\hat{\phi}\,,\\
&\Delta_{\omega_1} \Delta_{\omega_2} (\hat{\phi}\ast \hat{\psi}) = \Delta_{\omega_2} \Delta_{\omega_1} (\hat{\phi}\ast \hat{\psi})=-4\pi^2\,\delta\,. \label{eq:ExConv}
\end{align}
\end{Ex}

It is clear that if we want to find a suitable resummation procedure for a particular formal power series of interest these objects will play a crucial role, but before being able to apply this machinery we need to understand the following questions:
\begin{itemize}
\item What kind of generalisation to formal power series in $1/z$ do we expect? Trans-Series Section \ref{sec:Trans};
\item How do we compute in practice $\Delta_\omega$ ? Bridge equations Section \ref{sec:Bridge};
\item And finally how do we find the physical resummation procedure ? Median resummation and Stokes phenomenon Section \ref{sec:Median}.
\end{itemize}

\section{Intermezzo on Trans-Series}
\label{sec:Trans}

In this intermezzo we will introduce some basic concepts regarding the \textit{trans-series} expansion.
For a more complete overview of the subject we refer to \cite{Edgar:2008ga}. 
\begin{Def}
A \textit{Log-free trans-monomial} is a symbol of the form
\begin{equation}
\mathfrak{g}=z^a e^{T}
\end{equation}
with $a \in \mathbb{R}$ and $T$ is a purely large log-free trans-series.
\end{Def}
These trans-monomial are the building blocks of trans-series and they come with an order relation denoted by $\gg$ given by the relation
\begin{equation}
z^{a_1} e^{T_1} \gg z^{a_2} e^{T_2}
\end{equation}
if either $T_1 > T_2$ (where the symbol $>$ for trans-series will be defined shortly) or if $T_1=T_2$ and $a_1 > a_2$ as real numbers.
In example
\begin{equation*}
e^{e^z} \gg e^z \gg z^{-2} e^z \gg z^{10}\,.
\end{equation*}
\begin{Def}
A \textit{Log-free trans-series} is a formal sum of symbols
\begin{equation}
T = \sum_j c_j \mathfrak{g}_j
\end{equation}
where the coefficients $c_j \in \mathbb{R}$ and the $\mathfrak{g}_j$ are Log-free trans-monomial.
\end{Def}
The \textit{height} of a trans-monomial $z^a e^T$ is defined as the number of times we compose the formal exponential symbol, i.e. $z\,e^{e^z+z}$ has height $2$. Usually only finite height trans-monomial and trans-series are considered.

We just defined the trans-monomial using the notion of a trans-series and defined the trans-series starting from trans-monomials, in an Ouroboros manner. It is possible (see \cite{Edgar:2008ga}) to give a more precise definition of trans-monomials and trans-series in terms of Hahn series defined on the ordered abelian group of monomials but we will not need this level of sophistication.

We will say that the trans-series $T$ is \textit{purely large} if $\mathfrak{g}_j\gg 1$ for all the trans-monomial $\mathfrak{g}_j$ in $T$. Similarly a trans-series will be \textit{small} if $\mathfrak{g}_j \ll 1 $ for all $j$.
Alternatively we can call large term inifinite and small term infinitesimal since we are implicitly assuming the limit $z\to+\infty$.
A non-zero trans-series $T = \sum_j c_j \mathfrak{g}_j$ has a \textit{leading} term (also called \textit{dominance}) $dom(T)=c_0 \mathfrak{g}_0$ with the leading monomial (also called \textit{magnitude} of $T$) $mag(T)=\mathfrak{g}_0 \gg \mathfrak{g}_j$ for all the other terms present in $T$.

If the coefficient $c_0$ of the dominant term is positive, we say that the trans-series is positive and write $T>0$. In this way we can define an order relation between trans-series defined by $T>S$ iff $T-S > 0 $. Similarly if the $mag(T) \gg mag(S)$ we will write $T \gg S$, while if they have the same behaviour for $z \to\infty$, meaning that $dom(T)=dom(S)$, we will say that $T$ is \textit{asymptotic} to $S$, $T \sim S$. Note that only the zero trans-series can be asymptotic to $0$.

Trans-series inherit almost all standard properties of usual power series treated as formal sums.
In example differentiation of a trans-series is defined by the standard differentiation of trans-monomial
\begin{align}
&\mathfrak{g}'=\left( z^a e^T \right)' = a \,z^{a-1} e^T+z^a\, T' e^T\,, \\
&T'=\left(\sum_j c_j \mathfrak{g}_j\right)' = \sum_j c_j \mathfrak{g}_j'\,.
\end{align} 

A general trans-series is obtained by replacing for some $z$ inside a log-free trans-series the symbol $\log_m z$, with the identification
\begin{equation}
\log_m z =\log \circ \,... \circ \log z
\end{equation}
where we composed the logarithm $m\in \mathbb{N}$ times. The integer $m$ is called \textit{depth} of the trans-series.

Finite depth trans-series arise naturally when considering instanton contributions to physical observables. The instanton action plus perturbative corrections on top of that usually give rise to an height $1$ log-free trans-series, while the integration over the quasi-zero modes lead to the appearance of logarithmic corrections.
Hence in a generic theory with only one type of non-perturbative saddle points, a physical observable will take the form \cite{Dunne:2013ada,Dunne:2014bca}
\begin{equation}
E(z) = \sum_{n=0}^\infty \sum_{k=0}^{n-1} \left( z^\alpha\, e^{-S_0 z}\right)^n   \left[ \left(\log z\right)^k \,E^+_{n,k}(z) +\left(\log (-z)\right)^k \,E^-_{n,k}(z)\right]\,,
\label{eq:LogTrans}
\end{equation}
with $E^\pm_{n,k}(z)$ an height $0$, log-free trans-series, a.k.a. the asymptotic perturbative expansion around the $n$-instantons sector
\begin{equation}
E^\pm_{n,k}(z) = \sum_{p=0}^\infty c^\pm_{n,k,p}\, z^{-p-1}\,.
\end{equation}

Two comments are in order. Firstly we notice that the logarithms start appearing only at level $n=2$, what we would call two instantons sector. The reason is that in quantum mechanics \cite{Bogomolny:1980ur,ZinnJustin:1981dx,ZinnJustin:1982td} and quantum field theories \cite{Dunne:2012ae,Cherman:2014ofa}, the log sector is coming from the integration over quasi-zero modes. Quasi-zero modes are not exact zero modes, but nonetheless they are parametrically suppressed compared to genuine gaussian modes. The $n-1$ relative distances between $n$ different instantons are not exact zero modes because of intanton/(anti)instanton iteractions, when we integrate over these separations we will generate precisely between $0$ and $n-1$ logarithms, as suggested by the sum over $k$ in (\ref{eq:LogTrans}).

Secondly, it is striking that generic physical observables can be described by very easy trans-series, without having to use  logarithms with depth bigger than one, i.e. $\log (\log(z))$, or more involved exponential terms, i.e. $e^{e^{1/z}}$. A possible explanation might be traced back to the path integral formulation of the theory.  In the last Section we will see in a concrete, finite dimensional, example that the semiclassical decomposition of the functional integral as a sum over steepest descent contours would give rise to precisely only height $1$, depth $1$ trans-series. Unfortunately a full fledged path integral derivation of this result is still missing.

Hyperasymptotic expansions are extremely useful also in the context of linear and non-linear ODEs, where the connection with resurgence is well established, see \cite{Delabaere:2006ed} for the well known example of the Airy function.
In many cases we do not know an explicit solution to a given problem and we are forced to exploit asymptotic methods to get a feeling of how the actual solution might behave. In various interesting cases a simple power series expansion is not good enough and one has to use a trans-series expansion.
We will not discuss in details the hyperasymptotic expansion for ODEs, so we refer to the literature \cite{costin2008asymptotics} for a more detailed exposition of this interesting subject.
It is nonetheless instructive to analyse through a concrete example, based on \cite{Gair:2011mr}, how to implement this machinery of trans-series expansion in ODEs. 

\begin{Ex}
Let's study the non-linear ODE
\begin{equation}
y'(x) = \cos\left( \pi x\,y(x) \right)\,.
\label{Eq:NonLinearODE}
\end{equation}
While for many interesting physical problems \cite{Gair:2011mr}, a complete knowledge of the space of solutions is required, no explicit solution is actually known.

If we look for a solution going to $0$ for $x \gg 1 $, we can assume that it has an asymptotic expansion of the form
\begin{equation}
y(x) \sim \frac{a_0} {x} + \frac{a_1}{x^2} + O(x^{-3})\,.\label{eq:AsyODE}
\end{equation}
By plugging this ansatz in (\ref{Eq:NonLinearODE}), we see that $y'(x)$ vanishes for large $x$ if $ a_0 = n+1/2$ with $n\in\mathbb{Z}$.
After we fix the coefficient $a_0$ all the remaining coefficients are uniquely fixed in terms of $n$, i.e. $a_1=0,a_2=(-1)^n (n+1/2)/\pi$ and so on.
Something strange is going on here, we know that the solutions to this first order ODE should come with an arbitrary constant, i.e. $y(0)$, while instead it looks like we have a discrete set of solutions which asymptotically behave as $y(x)\sim 1/(2x),\,3/(2x),\, 5/(2x),...$, how does the initial condition $y(0)$ enter our asymptotic expansion?

We can solve numerically (\ref{Eq:NonLinearODE}), and as we vary the initial condition $y(0)=y_0$ the different solutions fall into disjoint classes, see Figure \ref{Fig:ODE}, with discrete asymptotic behaviour of the form $y(x) \sim (n+1/2)/x+...$ and $n$ even. From our previous perturbative expansion we were expecting solutions with asymptotic form $y(x) \sim (n+1/2)/x+...$ for all integers $n$, but from our numerics we find only behaviours like $1/(2x), 5/(2x), 9/(2x)$, so what happened to the solutions with $n$ odd?

\begin{figure}
\begin{center}
	\includegraphics[scale=0.3]{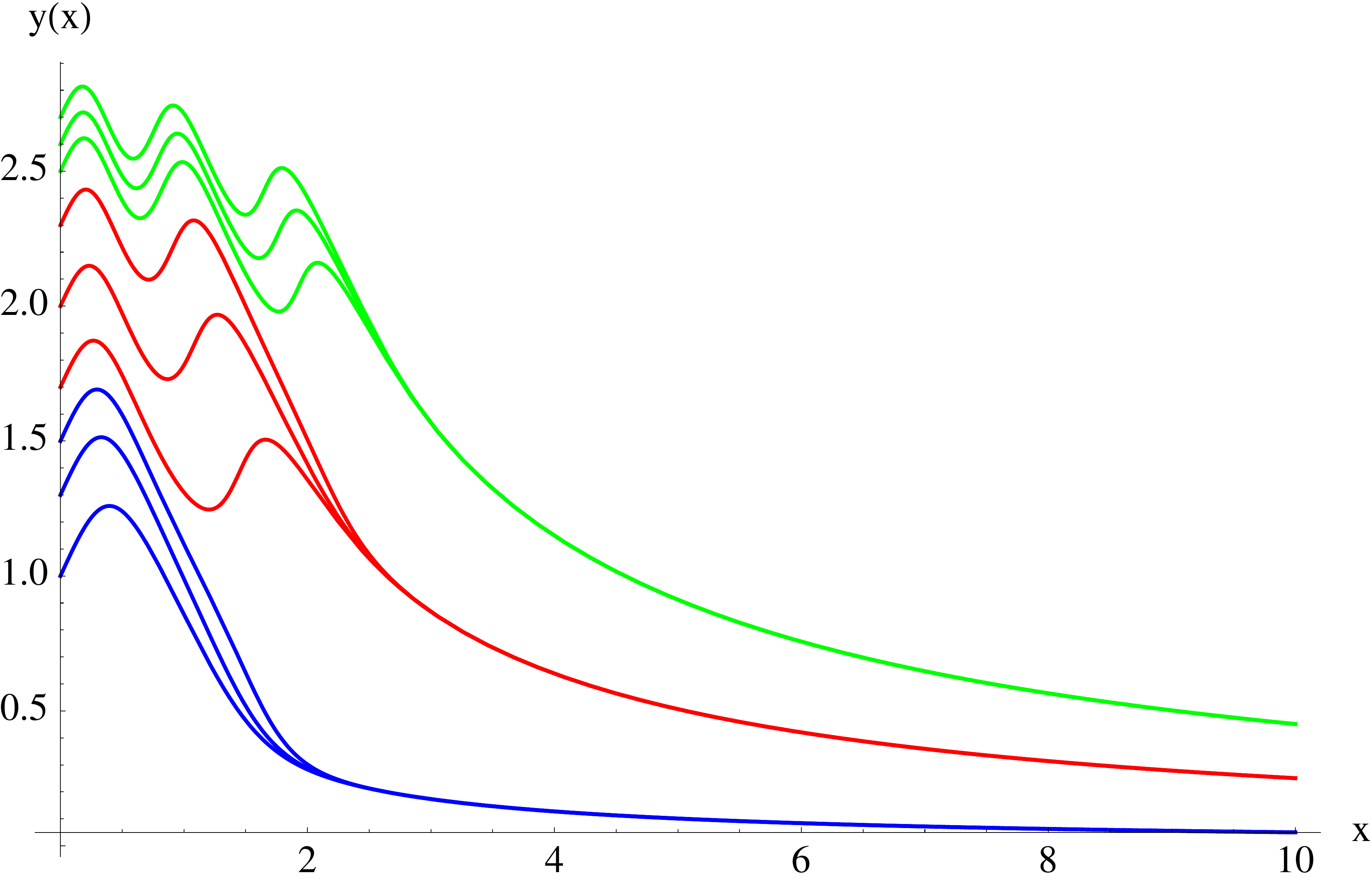}
	\caption{Numerical solutions with asymptotic behaviours of the form $1/(2x)$ (blue), $5/(2x)$ (red), $9/(2x)$ (green).}
	\label{Fig:ODE}
\end{center}
\end{figure}

It turns out that in order to find the solutions with asymptotic form $y(x) \sim (n+1/2)/x+...$ and $n$ odd, one has to give a precise (and unique) initial condition $y(0)=y_0^{(n)}$, all the solutions with initial data slightly off from this particular value, will fall either into the set with $n+1$ or $n-1$ (both even) asymptotic form.

To understand better this phenomenon let's assume that $y_1(x)$ and $y_2(x)$ are both solutions to (\ref{Eq:NonLinearODE}) with the exact same $a_0=(n+1/2)$, and $n\in \mathbb{Z}$.
Since all the higher orders terms $a_i$ are uniquely fixed once we fix $n$, tha asymptotic forms for $y_1$ and $y_2$ are precisely the same, which in particular means that $y_1-y_2$ is asymptotically smaller than any power of $1/x$!
Let's define $u(x)=y_1(x) - y_2(x)$, we know that its asymptotic form for large $x$ cannot be of the form $b_0/x+b_1/x^2+...$, but we also know that $y_1$ and $y_2$ differ from each other just because $y_1(0) \neq y_2(0)$, so $u$ cannot vanish identically.
From (\ref{Eq:NonLinearODE}), we can deduce the ODE satisfied by $u(x)$:
\begin{equation}
u'(x) = y_1'(x) -y_2'(x) = -2 \sin\left( \frac{\pi x\,(y_1(x)+y_2(x))}{2}\right) \,\sin \left(\frac{\pi x \,u(x)}{2}\right)\,,
\end{equation}
where we used some trigonometric identities to rewrite the difference of two cosines as product of sines.
We can use at this point the fact that $y_1(x)+y_2(x)\sim 2(n+1/2)/x$, valid for large $x$, and obtain 
\begin{equation}
u'(x)\sim (-1)^{n+1} 2 \sin \left(\frac{\pi x \,u(x)}{2}\right)\sim (-1)^{n+1} \,\pi x\,u(x)\,.
\end{equation}
At this point it is straightforward to obtain the asymptotic form for $u(x)$
\begin{equation}
u(x)=y_1(x)-y_2(x) \sim \mbox{Const.}\, e^{-(-1)^n\,\pi\,x^2/2}\,.
\end{equation}
This equation answers all our previous questions: firstly, the arbitrary constant that we were missing in the asymptotic expansion (\ref{eq:AsyODE}) was actually hiding in the hyperasymptotic part, and secondly we see that for $n$ even the difference between two solutions with different initial value, but in the same asymptotic class, is exponentially suppressed, while for $n$ odd they deviate from one another exponentially fast.
Clearly, for $n$ odd, we are not expecting the solution $u(x) \sim e^{+ \pi x^2}$ to be valid since we have assumed $u$ small when we expanded our ODE. All we know is that two such putative solutions $y_1,y_2$, with different initial conditions but belonging to the same asymptotic class with $n$ odd, will have to deviate from one another. This means that only for a precise initial value $y_0^{(n)}$, that can be computed numerically, we find the unique separatrix solution.

The use of a simple trans-series expansion allows us to get a complete understanding of the space of solutions for linear and non-linear ODEs, even (and especially) when analytic solutions are not known!
\end{Ex}

After this brief introduction to trans-series, we can go back to the main story of this work. We will now see how to include more generic trans-series in the context of resurgence and why including such terms is actually the crucial step in obtaining well defined physical observables.

\section{Trans-series Expansion and Bridge Equation}
\label{sec:Bridge}

As we have anticipated in Section \ref{sec:Stokes}, the perturbative power series expansion to our favourite physics or maths problem, is usually insufficient to recover the correct solution. Just by studying the analytic properties of the Borel transform of the perturbative series, we understand that resurgent symbols, i.e. non-perturbative contributions of the form $e^{-S\,z}$, have to be included to obtain a consistent formal solution.
This non-analytic, non-perturbative terms will be accompanied by a standard perturbative expansion on top of them, and in many cases we will also have to include logarithmic corrections (due to resonance in the case of Painlev\'e ODEs \cite{Garoufalidis:2010ya,Aniceto:2011nu,Schiappa:2013opa} or integration over quasi-zero modes for multi-instantons solutions \cite{Jentschura:2004jg}).
A general solution to our problem will eventually take the form of a sum of trans-monomial \
\begin{equation}
z^\alpha \,\log_m z\,e^{S(z)} \,\tilde{\phi}(z)\,
\end{equation}
with $S(z)$ possibly a trans-series itself and $\tilde{\phi}(z)$ a simple resurgent function.

Given the particular linear or non-linear problem to solve, the first step is understanding the type of trans-monomial that we have to use to obtain the complete solution.
For the toy model we will focus on, our ansatz solution will only contain height-$1$, log-free trans-monomial of the form
\begin{equation}
e^{-S_0\,n\,z} \, \tilde{\phi_n}(z)\,,
\end{equation}
where $S_0\in \mathbb{R}^+$ will be our instanton action and $\tilde{\phi_n}$ will be the perturbative expansion around the $n$-instantons solution.
This one parameter trans-series mimic a toy model in which we have only one type of non-perturbative configurations with real and positive action $S_0$, this is usually the case when the problem at hand depends on just one single boundary condition.
It is possible to obtain theories where one has multiple instantonic configurations with different actions $S_a,\,S_b,...$ \cite{Aniceto:2011nu}, and even complex valued action for the so called ghost-instantons \cite{Basar:2013eka} or more generically non-topological saddle points \cite{Cherman:2013yfa}.
For a more complete treatment of multi-parameter trans-series with the inclusion of logarithmic sectors we refer to the thorough works of Aniceto, Schiappa and Vonk \cite{Aniceto:2011nu,Aniceto:2013fka}.

We will give more explicit examples later on, but for the moment we assume that our perturbative formal solution give rise to a resurgent function $\tilde{\phi}_0(z)$, with singularities located at $S_0\,\mathbb{Z}^\ast$, for some instanton action $S_0\in\mathbb{R}$. This means that the Stokes automorphisms $\mathfrak{S}_{0}$ and $\mathfrak{S}_\pi$ will act non-trivially on $\tilde{\phi}_0$.
We are expecting some non-perturbative effect to modify our simple formal series ansatz turning it into the trans-series form
\begin{equation}
\Phi(z) = \sum_{n=0}^\infty \left(e^{- S_0\,z}\right)^n \,\tilde{\phi}_n(z)\,,
\end{equation}
where we can interpret the various simple resurgent functions $\tilde{\phi}_n$ as the perturbative contributions on top of the $n$-instanton configuration.
It is useful to introduce an additional complex parameter $\sigma$ to keep track of the resurgent symbols $e^{-S_0\,z}$, we will discuss then the one-parameter trans-series
\begin{equation}
\Phi(z,\sigma) = \sum_{n=0}^\infty \sigma^n\,\left(e^{- S_0\,z}\right)^n \,\tilde{\phi}_n(z)\label{eq:1ParamTS}\,.
\end{equation}

\begin{Ex}
For concreteness let's study a Riccati ODE
\begin{equation}
\frac{\partial \phi}{\partial z} -a\,\phi +\frac{1}{z^2} \phi^2=-\frac{1}{z}\,.\label{eq:Riccati}
\end{equation}
Without the non-linear term $\phi^2$ this equation is a simple generalisation of Euler's equation (\ref{eq:Euler})
\begin{equation}
\frac{\partial \psi}{\partial z} -a\,\psi =-\frac{1}{z}\,,
\end{equation}
 whose solution has the formal asymptotic form $\psi_0=\sum_{n\geq0} (-1)^n\,n!/ (a\,z)^{n+1} $ and its Borel transform takes the nice form $\mathcal{B}[\psi_0] = 1/ (a+\zeta)$, which is clearly a simple resurgent function with just one singularity at $\zeta = -a$.
The non linearity makes things more interesting: the solution $\psi_0$ gets modified
\begin{equation}
\phi_0 = \sum_{n=0}^\infty \frac{(-1)^n\,n!}{(a\,z)^{n+1}}\,c_n(a)=\frac{1}{ a\,z}-\frac{1}{(a\,z)^2} +\frac{2}{(a\,z)^3}- \frac{6}{(a\,z)^4}\left(1-\frac{a}{6}\right)+...\,,
\end{equation}
where the coefficient $c_n(a)$ are polynomials of degree $\floor{ n/3} $ in $a$, defined by the following recursion relation
\begin{align}
&\notag c_0 = c_1 = c_2 =1\,,\\
& c_n= c_{n-1} - a \sum_{l=0}^{n-3} \frac{(n-l-3)!\,l!}{n!} \,c_{n-l-3}\,c_l\,,\,\qquad n\geq 3\,.
\end{align}
It is possible to prove \cite{Delabaere:1999ef} that $\phi_0$ is of Gevrey-1 type and that its Borel transform $\mathcal{B}[\phi_0]$ has simple singularities for $\{-a,-2a,-3a,...\}$, hence, for this particular problem, the Riemann surface $\mathcal{R}$, defined in Section \ref{sec:Resurge}, is simply given by the universal covering of $\mathbb{C}\setminus\{-a,-2a,-3a,..\}$, and $\mathcal{B}[\phi_0]$ defines a simple resurgent function in $\widehat{\mbox{RES}}^{simp}\subset \delta\,\mathbb{C}\oplus\widehat{\mathcal{H}}(\mathcal{R})$. 

From the perturbative solution $\phi_0$, we can build a one parameter family, $\sigma\in\mathbb{C}$, of solutions
\begin{equation}
\Phi(z,\sigma;a) = \sum_{n=0}^\infty \sigma^n\,\left(e^{a\,z}\right)^n \,\phi_n(z)\,,\label{eq:TSRiccati}
\end{equation}
where $\Phi(z,\sigma;a)$ is an height one, log-free trans-series. Each $\phi_n$ is computed by identifying term by term the different powers of $\sigma$ when substituing in (\ref{eq:Riccati})
\begin{align}
&\frac{\partial \phi_1}{\partial z}+\frac{2}{z^2} \,\phi_0\,\phi_1=0\,,\\
&\frac{\partial \phi_2}{\partial z}+a\,\phi_2+\frac{2\phi_0\,\phi_2}{z^2}=-\frac{\phi_1^2}{z^2}\,.
\end{align}
Generically substituting a trans-series ansatz into a non-linear problem will give us a non-linear equation for $\phi_0$, that we will have to solve perturbatively, as we just did. The equation for $\phi_1$ will then be linear and homogeneous, while the one for the higher terms $\phi_{n\geq2}$ will be linear but inhomogeneous.
It turns out that for the normalization choice $\phi_1 = 1+O(z^{-1})$ there exists a unique formal solution to our Riccati equation (\ref{eq:Riccati}) of the trans-series form (\ref{eq:TSRiccati}) where each resurgent symbol $e^{n\,a\,z} \phi_n(z)$ gives us a simple resurgent function $\mathcal{B}[\phi_n]$ defined on the universal covering of $\mathbb{C}\setminus \{(n-1)a,(n-2)a,...,0,-a,-2a,...\}$. The Riemann surface $\mathcal{R}$ of Section \ref{sec:Resurge} for this problem is simply $\mathbb{C}\setminus\,a\,\mathbb{Z}$. 
\end{Ex}

So let's go back to our general discussion and assume that we have constructed our height one, log-free trans-series ansatz (\ref{eq:1ParamTS}) for the particular non-linear problem to solve.
As already stated in Section \ref{sec:Stokes}, the Stokes automorphisms along some singular direction (in this case either $\theta=0$ or $\theta=\pi$ since we assumed $S_0 \in \mathbb{R}$) is entirely captured by the Alien derivatives along that particular direction.
The problem is: we do not now how to compute for a generic trans-series (\ref{eq:1ParamTS}) its alien derivative, say at the singular point $\omega= k\,S_0$, for some $k\in\mathbb{Z}^\ast$.
This is where we need to relate the alien derivative to the standard derivative through some \textit{bridge equation}, which builds a bridge between alien calculus and standard differential calculus.

Suppose that $\Phi(z,\sigma)$ is the solution to some non-linear problem (i.e. finite difference equations for matrix models, Painlev\'e ODE for minimal strings or WKB for energy eigenvalues) in the variable $z$, then we know that
\begin{equation}
\left[ \partial_z, \dot{\Delta}_{k\,S_0}\right] = 0\,,
\end{equation}
which means that $ \dot{\Delta}_{k\,S_0}\Phi(z,\sigma) $ solves a linear homogeneous differential equation in $z$. Similarly, since $[\partial_z,\partial_\sigma]=0$, $\partial_\sigma \Phi(z,\sigma) $ solves exactly the same linear homogeneous problem (modulo some caveat on the initial data).
Since $ \dot{\Delta}_{k\,S_0}\Phi(z,\sigma)$ and $\partial_\sigma \Phi(z,\sigma) $ are solutions to the same linear homogeneous ODE, say of order 1 for example, they must be proportional to each others
\begin{equation}
\dot{\Delta}_{k\,S_0}\Phi(z,\sigma) = A_k(\sigma) \,\partial_\sigma \Phi(z,\sigma)\,.\label{eq:Bridge}
\end{equation}
This is called \textit{Ecalle's Bridge equation}, it gives us a bridge to relate the Alien derivative to the usual derivative in the trans-series parameter $\sigma$.

\begin{Ex}
Let's go back to Riccati ODE (\ref{eq:Riccati}) and our trans-series formal solution $\Phi(z,\sigma;a)$ (\ref{eq:TSRiccati}).
The trans-series $\Phi(z,\sigma;a)$ solves
\begin{equation*}
\frac{\partial \Phi}{\partial z} -a\,\Phi +\frac{1}{z^2} \Phi^2=-\frac{1}{z}\,,
\end{equation*}
so let's apply $\dot{\Delta}_{na}$, with $n\in \mathbb{Z}^\ast$, on both sides
\begin{equation}
\partial_z \, \left(\dot{\Delta}_{na} \Phi \right)-a\,\dot{\Delta}_{na} \Phi +\frac{2\,\Phi}{z^2} \dot{\Delta}_{na} \Phi = 0\,,
\end{equation}
where we used that $\dot{\Delta}$ commutes with $\partial_z$ together with the fact that $\Delta$ acts as a derivation (i.e. Jacobi holds). Note that $\dot{\Delta}_{na} 1/z=0$ and $\dot{\Delta}_{na} 1/z^2=0$ since their Borel transform are entire function along the real line.
If we apply now $\partial_\sigma$ on both sides of the Riccati equation we get
\begin{equation}
\partial_z \left(\partial_\sigma \Phi\right) -a\,\partial_\sigma \Phi +\frac{2\,\Phi}{z^2} \partial_\sigma \Phi = 0\,.
\end{equation}
As anticipated $\dot{\Delta}_{na}\Phi$ and $\partial_\sigma \Phi$ are both solutions to the same homogeneous, order one equation in $\partial_z$, hence they must be proportional
\begin{equation}
\dot{\Delta}_{na} \Phi (z,\sigma;a) = A_n(\sigma;a)\,\partial_\sigma \Phi(z,\sigma;a)\,,
\end{equation}
which is precisely Ecalle's Bridge equation.
\end{Ex}

Ecalle's Bridge equation is the crucial missing piece of the puzzle, with this equation we can relate the mysterious alien derivative to standard calculus. This equation tells us that, at all singular points, the alien derivative gives back the original asymptotic expansion, hence the name \textit{resurgence}.
Let's investigate further the Bridge equation (\ref{eq:Bridge}). Focusing on the LHS we get
\begin{equation}
\dot{\Delta}_{kS_0} \Phi = \sum_{n=0}^\infty \sigma^n e^{-(n+k)S_0\,z} \Delta_{kS_0} \tilde{\phi}_n\,,
\label{eq:br1}
\end{equation}
while on the RHS
\begin{equation}
A_k(\sigma) \,\partial_\sigma \Phi(z,\sigma) = \sum_{n=0}^\infty A_k(\sigma) n\,\sigma^{n-1} e^{-n S_0\,z} \tilde{\phi}_n\,.
\label{eq:br2}
\end{equation}
We have to match term by term on the two sides, with exactly the same power of $\sigma^m$ and the same resurgent symbol $e^{-n\,S_0\,z}$. Since $\dot{\Delta}_{kS_0} \Phi$ contains only positive powers of $\sigma$ we can assume that
\begin{equation}
A_k (\sigma) =\sum_{m=0}^\infty A_{k,m} \sigma^m\,,
\end{equation}
for some complex numbers $A_{k,m}$.
Furthermore in $\Phi$, each resurgent symbol $e^{-n\,S_0\,z}$ is accompanied by precisely $\sigma^n$ and since $\dot{\Delta}_{kS_0}$ introduces an additional $e^{-kS_0\,z}$, it means that, to restore the degree between $\sigma$ and $e^{-S_0 \,z}$, we must have $A_k(\sigma) =A_k \, \sigma^{1-k}$.

By matching each term in (\ref{eq:br1}) with the terms in (\ref{eq:br2}), with exactly the same power of $\sigma$ and the same resurgent symbol $e^{-n\,S_0\,z}$, we obtain the set of equations
\begin{align}
&\Delta_{kS_0} \tilde{\phi}_n = 0\,,\,\qquad\qquad \qquad \qquad k> 1\,,\label{eq:BridgeComponent1}\\
&\Delta_{kS_0} \tilde{\phi}_n =  A_k\,(n+k)\,\tilde{\phi}_{n+k}\,,\qquad k\leq 1\,,\label{eq:BridgeComponent}
\end{align}
with the definition $\tilde{\phi}_n=0$ for $n < 0$, and where the complex constants $A_k \in\mathbb{C}$ are called \textit{holomorphic or analytic invariants} of the problem. In principle we would know all the alien derivatives if we knew all the $A_1,A_0,A_{-1},...$, needless to say the various $A_k$ are really hard to compute.

\textbf{Remark:} Note that the vanishing of $\Delta_{kS_0} \tilde{\phi}_n$ for $k >1$ does not imply that $k S_0$ is a regular point! As we have already seen before in (\ref{eq:ExConv}) the singular behaviour is known once we know \textit{all} the multiple alien derivatives, in example we have $\Delta_{2 S_0} \tilde{\phi}_0 =0 $ while $\Delta_{S_0} \Delta_{S_0} \tilde{\phi}_0 =2\,A_1^2\,\tilde{\phi}_2$.
The singular behaviour close to $2S_0$ of what we would call the perturbative series $\tilde{\phi}_0$ is entirely captured (and vice versa) by the perturbative expansion around the 2-instantons contribution. Once again the perturbative series surges up, or resurges, from the non-perturbative physics, furthermore, since $\Delta_{2 S_0} \tilde{\phi}_0 =0 $ while $\Delta_{S_0}^2\tilde{\phi}_0 \neq 0$, we know that this new singular point is not associated with a new non-perturbative object with action $2S_0$, but rather it arises from a multi-instanton saddle.

From the definition of Alien derivative (\ref{eq:AlienEasy}), we see that the Bridge equations (\ref{eq:BridgeComponent1})-(\ref{eq:BridgeComponent}) tell us that, close to the singular point $kS_0 $, the singular behaviour of the simple resurgent function $\tilde{\phi}_n$ is entirely governed by $\tilde{\phi}_{n+k}$ since (passing to the convolutive model now)
\begin{equation}
 \mathcal{B}[\tilde{\phi}_n] (\zeta+k\,S_0) \sim A_k (n+k)\,\mathcal{B}[\tilde{\phi}_{n+k}](\zeta) \,\log \zeta / 2 \pi i\,.
\end{equation}
We were a little bit too sketchy here, as we know, the precise definition of Alien derivative (\ref{eq:AlienPlus}) is more complicated than that, but the main point still remains: the singular part of $\tilde{\phi}_n$ at $k S_0$ is entirely captured by $\tilde{\phi}_{n+k}$.

The Bridge equations (\ref{eq:BridgeComponent1})-(\ref{eq:BridgeComponent}) not only allow us to reconstruct the entire behaviour of our trans-series close to a singular point but they also make manifest the appearance of the Stokes phenomena along the singular lines $\theta=0$ and $\pi$.
To see that, let's go back to the expression for the Stokes automorphism in term of Alien derivative (\ref{eq:expAlien}) and specialise it to the singular direction $\theta=0$ \footnote{For the direction $\theta=\pi$ the situation is a little bit more involved but as we will show later on the end results will be the same}
\begin{equation}
\mathfrak{S}_0 = \exp \left( \sum_{k=1}^\infty e^{-k\,S_0\,z} \Delta_{kS_0} \right)\,.
\end{equation}
Given our trans-series ansatz and the Bridge equations (\ref{eq:BridgeComponent1})-(\ref{eq:BridgeComponent}), we already know that $\Delta_{kS_0} \tilde{\phi}_n = 0$ for all $n$ as soon as $k>1$, for this reason the above equation simplifies drastically to
\begin{equation}
\mathfrak{S}_0 = 1+ e^{-S_0\,z} \Delta_{S_0} + \frac{1}{2} e^{-2S_0\,z} \Delta^2_{S_0}+...\,.
\end{equation}
It is easy to compute multiple alien derivatives just by iterating
\begin{equation}
\Delta_{S_0} \tilde{\phi}_n = A_1\,(n+1)\,\tilde{\phi}_{n+1}\,,
\end{equation}
so that
\begin{equation}
\Delta_{S_0}^k \tilde{\phi}_n = A^k_1\,(n)_k\,\tilde{\phi}_{n+k}\,,
\end{equation}
where we used the Pochhammer symbol $(n)_k = \prod_{i=1}^k (n+i)$.
We have now all the ingredients to compute the Stokes automorphism along the positive real line
\begin{equation}
\mathfrak{S}_0 \tilde{\phi}_n = \sum_{k=0}^\infty \frac{1}{k!} \Delta_{S_0}^k \tilde{\phi}_n = \sum_{k=0}^\infty {n+k\choose n} A_1^k\,e^{-k\,S_0\,z}\,\tilde{\phi}_{n+k}\label{eq:StokesPartial}\,.
\end{equation}

We can use the definition (\ref{eq:Stokes}) of the Stokes automorphism to relate the two sectorial sums above, $\mathcal{S}_{0^+}$, and below, $\mathcal{S}_{0^-}$, the positive real axis
\begin{align}
\mathcal{S}_{0^+} \Phi(z,\sigma) &\notag =  \mathcal{S}_{0^-} \circ \mathfrak{S}_0  \Phi(z,\sigma) \\
&\notag= \mathcal{S}_{0^-} \left(1+e^{-S_0\,z} \Delta_{S_0}+... \right) \left(\sum_{n=0}^\infty \sigma^n\,e^{-nS_0\,z} \tilde{\phi}_n \right)\,\\
& = \mathcal{S}_{0^-} \left[ \sum_{n=0}^\infty \sigma^n e^{-nS_0\,z} \left( \sum_{k=0}^\infty {n+k\choose n} A_1^k\,e^{-k\,S_0\,z}\,\tilde{\phi}_{n+k} \right)\right]\,.
\end{align}
Note that even if the Stokes automorphism, when applied to $\tilde{\phi}_n$, generates an infinite sum (\ref{eq:StokesPartial}), nonetheless each resurgent symbols $e^{- mS_0\,z}$ in $\mathfrak{S}_0 \Phi$ receives contributions only from a finite number of terms, precisely from $n,k \in \mathbb{N}$ such that $n+k = m$. We can thus change variables in the sum from $n,k \in \mathbb{N}$ to $m=n+k\in \mathbb{N}$ and $p \in \{0,1,..., m\}$ and arrive at
\begin{align}
\mathcal{S}_{0^+} \Phi(z,\sigma) &= \mathcal{S}_{0^-} \left[\sum_{m=0}^\infty  e^{- mS_0\,z}\,\tilde{\phi}_m  \left( \sum_{p=0}^m {m \choose p} \sigma^{m-p} \,A_1^p \right)\right]\\
& =  \mathcal{S}_{0^-} \left[ \sum_{m=0}^\infty  e^{- mS_0\,z}\,\tilde{\phi}_m \left(\sigma+A_1\right)^m\right]\,,
\end{align}
comparing this to our original expansion (\ref{eq:1ParamTS}) we have finally found
\begin{equation}
\mathcal{S}_{0^+} \Phi(z,\sigma) = \mathcal{S}_{0^-} \Phi(z,\sigma+A_1)\,.\label{eq:StokesPheno}
\end{equation}
We could have obtained the same result directly from the original Bridge equation written in terms of $\Phi$
\begin{equation}
\dot{\Delta}_{k S_0} \Phi(z,\sigma)= A_k \,\sigma^{1-k}\, \frac{\partial \Phi}{\partial \sigma}\,,\label{eq:Bridge2}
\end{equation}
valid for all $k \leq 1$ different from zero, and the Stokes automorphism along $\theta=0$ becomes
\begin{equation}
\mathfrak{S}_0 \Phi(z,\sigma) = \exp\left(\dot{\Delta}_{-S_0} \right) \Phi(z,\sigma) = \exp\left(A_1\frac{\partial}{  \partial\sigma} \right)\Phi(z,\sigma)= \Phi(z,\sigma+A_1)\,.\label{eq:StokesZero}
\end{equation}

The equation just obtain is a beautiful summary of all our alien calculus journey: along a singular direction, say the positive real line, the resummed series when $\theta=0^+$ can be obtained by the resummed series for $\theta=0^-$ plus a jump in the trans-series parameter $\sigma$ exactly equal to Ecalle's holomorphic invariant $A_1$.
The Stokes phenomenon is encoded perfectly in the trans-series analysis of the Bridge equations, the only thing we are left to understand is how to define a non-ambiguous, unique (and possibly real, depending on the case) sum for our trans-series across a singular direction. This will be the aim of the next Section.

For completeness, let's analyse what happens to the trans-series ansatz and the Stokes automorphism along the singular direction $\theta = \pi$.
The Bridge equations (\ref{eq:BridgeComponent1})-(\ref{eq:BridgeComponent}) tell us that all the Alien derivatives $\Delta_{-kS_0}$, with $k=1,2,..$, will act non-trivially in $\mathfrak{S}_{\pi}$
\begin{equation}
\mathfrak{S}_\pi = \exp \left(\sum_{k=1}^\infty e^{ k S_0 \,z} \Delta_{-k S_0}\right)=1+e^{S_0\,z} \Delta_{-S_0}+e^{2S_0\,z}\left(\Delta_{-2S_0}+\frac{1}{2} \Delta_{-S_0}^2\right)+...\,.\label{eq:StokesPi}
\end{equation}
We have to compute the action of multiple alien derivatives on each simple resurgent function $\tilde{\phi}_n$ since the contributions to each resurgent symbol $e^{k S_0\,z}$ in $\mathfrak{S}_\pi$ come from
\begin{equation}
\Delta_{-k_1 S_0}...\Delta_{-k_N S_0} \tilde{\phi}_n
\end{equation} 
where the $\{k_i\}$ are all the possible integer partitions of $k=k_1+...+k_N,$ with $k_i \geq 1$.
Note as well that these are ordered partitions since the Alien derivatives at different points do not commute
\begin{equation}
[\Delta_{-k_1 S_0},\Delta_{-k_2 S_0}] \tilde{\phi}_n = A_{-k_1} \,A_{-k_2}\,(k_1-k_2)\,(n-k_1-k_2) \,\tilde{\phi}_{n-k_1-k_2}\,.
\end{equation}
Furthermore, from (\ref{eq:BridgeComponent1})-(\ref{eq:BridgeComponent}), we deduce that the infinite sum in $\mathfrak{S}_{\pi} \tilde{\phi}_n$ is actually a finite sum (contrary to the $\theta=0$ case) since as soon as we reach the level $e^{-nS_0\,z}$ we will have to compute some Alien derivative of the form
\begin{equation}
\Delta_{-k_1 S_0}...\Delta_{-k_N S_0} \tilde{\phi}_n\,,\qquad\qquad n= \sum_{i=1}^N k_i\,,
\end{equation}
which are all vanishing, together with each subsequent application of the alien derivative operator.
The generic iteration of multiple derivatives gives us
\begin{equation}
\prod_{i=1}^N \Delta_{-k_{(N+1-i)} S_0} \,\tilde{\phi}_n = \prod_{i=1}^N A_{-k_i} \cdot \prod_{i=1}^N \left(n-\sum_{j=1}^i k_j \right)\,\tilde{\phi}_{n-\sum_i k_i}\,,
\end{equation}
which clearly vanishes as soon as $\sum_{i=1}^N k_i \geq n$.

It is possible to obtain an analytic expression for $\mathfrak{S}_\pi \tilde{\phi}_n$ but it is not particularly illuminating \cite{Aniceto:2011nu,Aniceto:2013fka}, the important point to keep in mind is that along all the singular directions, the action of the Stokes automorphism on $\Phi(z,\sigma)$ can be recast in term of a differential operator acting on the trans-series parameter $\sigma$, giving rise to the Stokes phenomenon.
From equation (\ref{eq:Bridge2}) and the expression (\ref{eq:StokesPi}) for $\mathfrak{S}_\pi$ written in terms of alien derivatives we get
\begin{equation}
\mathfrak{S}_\pi \Phi(z,\sigma) = \exp\left(\sum_{k=1}^\infty \dot{\Delta}_{-kS_0} \right) \Phi(z,\sigma) = \exp\left(
\sum_{k=1}^\infty A_{-k} \,\sigma^{k+1} \frac{\partial}{\partial \sigma}
\right) \Phi(z,\sigma)\,.\label{eq:StokesPi}
\end{equation}

To get a feeling on how $\mathfrak{S}_\pi$ acts on $\Phi(z,\sigma)$, we can assume for the moment, that all the holomorphic invariants are vanishing except one, say $A_{-k}\neq 0$.
In this situation, the Stokes automorphism will simply be $\mathfrak{S}_\pi=\exp(A_{-k} \sigma^{k+1} \,\partial/ \partial\sigma)$, and its action on $\Phi(z,\sigma)$ is a simple translation of an associated trans-series parameter $\sigma^{-k}\rightarrow \sigma^{-k}-k\,A_{-k}$. This means that in this particular case where $A_{-k}$ is the only non-zero holomorphic invariant, the Stokes phenomenon along $\theta= \pi$ takes the form
\begin{equation}
\mathfrak{S}_\pi \Phi(z, \sigma) = \Phi(z, (\sigma^{-k}-k A_{-k})^{-1/k})\,,
\end{equation}
a generalisation of the $\theta=0$ case (\ref{eq:StokesZero}). Clearly when all the $A_{-k} $ are non vanishing the Stokes automorphism will be much more complicated, and given by (\ref{eq:StokesPi}).

Before concluding this Section, as a concrete example of what just discussed, we can study the case in which the trans-series contains only two terms, namely
\begin{equation}
F(z,\sigma_0,\sigma_1) = \sigma_0 F_0(z) + \sigma_1 F_1(z)\,,
\end{equation}
where the trans-monomials $F_l(z)$, with $l=0,1$, take the form
\begin{equation}
F_l (z) = e^{- M_l \,z} \Phi_l (z)  = e^{-M_l \,z} \sum_{n=0}^\infty a_n^{(l)} \,z^{-n-1}\,.\label{eq:2terms}
\end{equation}
For simplicity we will work with $M_0= 0$, which we will call the perturbative vacuum, and $M_1 =  M \in\mathbb{R}^+$, which we will call the NP-saddle, or instanton sector.

In this particular example, the only possible singular directions in the Borel plane will be $\theta=0$ and $\theta=\pi$. To compute the Stokes automorphism across these two singular directions we will need the Bridge equation (\ref{eq:Bridge}), which in this case takes the form
\begin{equation}
{\dot{\Delta}}_\omega F (z,\sigma_0,\sigma_1) = \sum_{l=0}^1 A_{\omega}^{[l]} (\sigma_0, \sigma_1)\frac{\partial F (z,\sigma_0,\sigma_1)}{\partial \sigma_l}\,,\label{eq:2TermBridge}
\end{equation}
where the undetermined functions $A_{\omega}^{[l]} (\sigma_0, \sigma_1)$ are related to the Stokes constants (analytic invariants).
We can Taylor expand these unknown functions
\begin{equation}
A_{\omega}^{[l]} (\sigma_0, \sigma_1) = \sum_{k,m\geq 0 } A^{[l]\,(k,m)}_{\omega} \sigma_0^k \, \sigma_1^m\,,
\end{equation}
where the complex numbers $A^{[l]\,(k,m)}_{\omega}$ are precisely the Stokes constants, non vanishing only for very few particular values of $\omega,l,k,m$.

We can expand the l.h.s. of (\ref{eq:2TermBridge}) 
\begin{equation}
{\dot{\Delta}}_\omega F (z,\sigma_0,\sigma_1) = \sum_{l=0}^1 \sigma_l \, e^{-(M_l+\omega) z} \Delta_\omega \Phi_l(z)\,,
\end{equation}
and substitute the Taylor expansion in the r.h.s. of (\ref{eq:2TermBridge}) to get
\begin{equation}
\sum_{l=0}^1 \sigma_l \, e^{-(M_l+\omega) z} \Delta_\omega \Phi_l(z) = \sum_{i=0}^1 \sum_{k,m\geq 0 } A^{[i]\,(k,m)}_\omega \sigma_0^k\sigma_1^m e^{-M_i z} \Phi_i(z)\label{eq:2BridgeExp}\,.
\end{equation}
The crucial point behind the trans-series expansion for the Bridge equation is that by matching equal powers of $\sigma_0,\sigma_1$ and $e^{-z}$, the allowed non vanishing Stokes constants will be enormously constrained.
In particular for this two parameters trans-series the only allowed constants are a subset of $T = \{ A^{[i]\,(1,0)},A^{[i]\,(0,1)}\}$, 
and these constants can be non-zero if and only if $M_l + \omega = M_i$ for some $\omega,l,i$.

We can specialise (\ref{eq:2BridgeExp}) to the singular direction $\theta=0$ for which we get
\begin{align}
\Delta_M \Phi_0 (z) &= A_M^{[1]\,(1,0) } \Phi_1(z) \,,\\
\Delta_M \Phi_1(z) & = 0 \,,
\end{align}
with all the other alien derivatives vanishing for all $\omega \in \mathbb{R}^+$ and $\omega \neq M$.
Similarly for the singular direction $\theta = \pi$, equation (\ref{eq:2BridgeExp}) becomes
\begin{align}
\Delta_{-M} \Phi_0 (z) &=0  \,,\\
\Delta_{-M} \Phi_1(z) & =  A_{-M}^{[0]\,(0,1) } \Phi_0(z) \,,
\end{align}
and once again all the other alien derivatives are vanishing for all $\omega \in \mathbb{R}^-$ and $\omega \neq- M$.
By renaming the only non vanishing Stokes constants $A_M^{[1]\,(1,0) }= A_M$ and $A_{-M}^{[0]\,(0,1) }= A_{-M}$, we can rewrite the entire resurgence algebra for this two-terms trans-series in the form
\begin{align}
&\notag \Delta_{M} \Phi_0(z) = A_M\,\Phi_1(z) \,,\qquad\qquad \,\Delta_{-M}\Phi_0(z) = 0\,, \\
&\label{eq:2TermResAlgebra}\Delta_M \Phi_1(z) = 0\,,\qquad\qquad\qquad \qquad\Delta_{-M} \Phi_1(z) = A_{-M} \, \Phi_0(z)\,.
\end{align}

Thanks to the above equations, the Stokes automorphism along $\theta = 0$ can be explicitly written as
\begin{equation}
\mathfrak{S}_0 = \exp \left( \sum_{\omega} e^{-\omega z} \Delta_\omega \right)=1+e^{-M z} \Delta_M\,,
\end{equation}
so that 
\begin{align}
\mathfrak{S}_0 \Phi_0 (z) &\notag = \Phi_0(z) + A_M\, e^{-M z} \Phi_1(z) \,,\\
\mathfrak{S}_0 \Phi_1(z) & = \Phi_1(z)\,.\label{eq:Stokes2plus}
\end{align}
This means that, as we approach the singular direction $\theta = 0$, $\Phi_1$ makes no jump while the jump of $\Phi_0$ is entirely dictated by $\Phi_1$.
For the full trans-series the Stokes automorphism along this direction is given by
\begin{align}
\mathfrak{S}_0 F(z,\sigma_0,\sigma_1) &\notag= \mathfrak{S}_0 \left( \sigma_0 \Phi_0(z) + \sigma_1 e^{-M z} \Phi_1(z)\right)\,\\
&\notag = \sigma_0 \left(\Phi_0(z) + A_M e^{- Mz} \Phi_1(z)\right)+\sigma_1 e^{-Mz} \Phi_1(z)\,,\\
&\notag = \sigma_0 \Phi_0(z) + \left(\sigma_1+\sigma_0 A_M\right) e^{-Mz} \Phi_1(z)\,,\\
& \label{eq:StokesZero2terms}= F(z, \sigma_0 ,\sigma_1 + A_M \sigma_0)\,.
\end{align}
Had we started with perturbation theory alone $F(z, \sigma_0=1,\sigma_1 =0)= \Phi_0(z)$, the Stokes automorphism along the singular direction $\theta$ would have generated for us a second recessive term 
\begin{equation}
\mathfrak{S}_0 F(z,1,0) = F(z,1,A_M)=\Phi_0(z)+A_M\,e^{-M z}\Phi_1(z)\,.
\end{equation}

In a similar manner, the Stokes automorphism along the singular direction $\theta = \pi$ is given by
\begin{equation}
\mathfrak{S}_\pi = \exp \left( \sum_{\omega} e^{-\omega z} \Delta_\omega \right)=1+e^{ + M z} \Delta_{-M}\,,
\end{equation}
so that 
\begin{align}
\mathfrak{S}_\pi \Phi_0 (z) &\notag  = \Phi_0(z) \,,\\
\mathfrak{S}_\pi \Phi_1(z) & =  \Phi_1(z)+ A_{-M}\,e^{+ M z}  \Phi_0(z)\,.\label{eq:Stokes2minus}
\end{align}
The roles are now inverted, $\Phi_0$ makes no jump along the negative real line while the entire jump of $\Phi_1$ is dictated by $\Phi_0$.
The Stokes automorphism along the negative real line for the two terms trans-series becomes
\begin{equation}
\mathfrak{S}_\pi F(z,\sigma_0,\sigma_1) = F(z,\sigma_0 + A_{-M} \sigma_1,\sigma_1)\,.
 \label{eq:StokesPi2terms}
\end{equation}

This complete knowledge of the Stokes automorphism allows us to study the large oder behaviour of the perturbative (and non-perturbative) coefficients $a_n^{(l)}$ in (\ref{eq:2terms}).
By Cauchy theorem\footnote{Note that generically we would get contributions coming from all the discontinuities across all the singular directions and from all the residues at simple poles in the Borel plane.} we know that
\begin{equation}
F(z) = \frac{1}{2\pi i} \oint \frac{F(\omega)}{\omega-z}=\frac{1}{2\pi i} \int_0^\infty d\omega \frac{Disc_0 F(\omega)}{\omega-z} +\frac{1}{2\pi i} \int_0^{-\infty} d\omega \frac{Disc_\pi F(\omega)}{\omega-z}\,,
\end{equation}
and by expanding for $z\to \infty$
\begin{equation}
\frac{1}{\omega-z} = - \sum_{n=0}^\infty \omega^n\,z^{-n-1}\,,
\end{equation}
we get
\begin{equation}
F_n \sim -\frac{1}{2\pi i} \int_0^\infty d\omega\,\omega^{n} Disc_0 F(\omega) -\frac{1}{2\pi i} \int_0^{-\infty} d\omega \,\omega^n Disc_\pi F(\omega)\,,\label{eq:Largeorder}
\end{equation}
where we schematically wrote $F(z) \sim \sum_{n\geq 0} F_n z^{-n-1}$.

We can specialise the above equations for the large orders behaviour of our perturbative expansion
\begin{equation}
\Phi_0(z) = \sum_{n=0}^\infty a_n^{(0)} z^{-n-1}\,,
\end{equation}
and thanks to (\ref{eq:Stokes2plus})-(\ref{eq:Stokes2minus}), we know the full discontinuities
\begin{align}
Disc_0 \Phi_0(z) &= \left( \mbox{Id}-\mathfrak{S}_0 \right) \Phi_0(z) = - A_M e^{-M z} \Phi_1(z)\,,\\
Disc_\pi \Phi_0(z) & =  \left( \mbox{Id}-\mathfrak{S}_\pi \right) \Phi_0(z) =0\,.
\end{align}
In (\ref{eq:Largeorder}) only the first term contributes and the large orders behaviour of the perturbative expansion is entirely controlled by the lower orders of the NP-saddle expansion \cite{berry1993unfolding}
\begin{align}
a_n^{(0)} &\notag \sim \frac{A_M }{2\pi  i} \sum_{ k\geq 0 } a_k^{(1)} \int_0^\infty d\omega \omega^{n-k} e^{- M \omega}\\
&\notag \sim \frac{A_M }{2\pi i} \sum_{k\geq 0} a_k^{(1)} \frac{\Gamma(n+1-k)}{ M^{n+1-k}}\\
&\label{eq:HigherOrderP} \sim \frac{A_M }{2\pi i} \frac{n!}{M^{n+1}}\left( a_0^{(1)} + a_1^{(1)} \frac{M}{n}+a_2^{(1)} \frac{M^2}{n(n-1)}+...\right)\,.
\end{align} 
Note that since $M>0$ these coefficients are non-alternating in sign for $n$ large enough.

The story for the large orders of $\Phi_1$ can be repeated verbatim.
Again using (\ref{eq:Stokes2plus})-(\ref{eq:Stokes2minus}), the discontinuities are
\begin{align}
Disc_0 \Phi_1(z) &= \left( \mbox{Id}-\mathfrak{S}_0 \right) \Phi_1(z) = 0\,,\\
Disc_\pi \Phi_1(z) & =  \left( \mbox{Id}-\mathfrak{S}_\pi \right) \Phi_1(z) =- A_{-M} e^{+ Mz} \Phi_0(z)\,.
\end{align}
The large orders behaviour for the NP-saddle perturbation theory is entirely captured by the lower orders coefficients of the perturbative vacuum
\begin{align}
a_n^{(1)} &\notag \sim \frac{A_{-M}}{2\pi i} \sum_{k\geq 0} a_k^{(0)} \frac{\Gamma(n+1-k)}{(-M)^{n+1-k}}\\
&\label{eq:HigherOrderNP} \sim \frac{A_{-M}}{2\pi i}\frac{n!}{(-M)^{n+1}} \left( a_0^{(0)} - a_1^{(0)} \frac{M}{n} + a_2^{(0)} \frac{M^2}{n(n-1)}+...\right)\,.
\end{align}
Note that, thanks to the $(-M)^{n+1}$ term, the NP-saddle perturbative coefficients are eventually alternating in sign, in constrast to the vacuum coefficients $a_n^{(0)}$. 

\textbf{Remark.} It is worth emphasizing again in words what we just found: the large orders coefficients of the perturbative expansion do contain explicitly the lower orders coefficients of the NP saddle expansion and vice versa!

In the next Section we will construct a toy model for path integral calculations in which we will be able to check explicitly how the lower orders non-perturbative coefficients are encoded and buried in the higher orders coefficients of the perturbative series.

\section{Median Resummation and Cancellation of Ambiguities}
\label{sec:Median}

In this final Section we will put everything together and show how a trans-series expansion for physical observables, obtained from a path integral, will automatically be free from ambiguities coming from the non-Borel summability, along the positive real line, of the perturbative series.

To be concrete and closest as possible to an actual path integral calculation, we can focus on a particular dimensional reduction, down to $0$ dimensions, of the $2d$ sine-Gordon model\footnote{See also \cite{Cherman:2014xia} for a discussion on the monodromies properties of the model.}, see \cite{Cherman:2014ofa}. The partition function is given by
\begin{align}
Z(\lambda) &\notag=
\frac{1}{\sqrt{4\lambda}} \int_{-\pi}^\pi dy \,e^{\frac{1}{4\lambda} (\cos y -1)}\\
&\notag=\frac{1}{\sqrt{\lambda}} \int_{-\pi/2}^{\pi/2} dx\,e^{-\frac{1}{2\lambda} \sin^2 x}\\
& = \frac{\pi}{\sqrt{\lambda}} e^{-\frac{1}{4\lambda}} I_0 \left(\frac{1}{4\lambda}\right)\,,
\label{eq:Z2terms}
\end{align}
where $I_q$ is the modified Bessel function of the first kind with index $q$.
We can also consider an other observable, related to the expectation value of the operator $e^{ i q y}$, given by
\begin{align}
\mathcal{O}_q(\lambda) &\notag=
\frac{1}{\sqrt{4\lambda}} \int_{-\pi}^\pi dy \,e^{\frac{1}{4\lambda} (\cos y -1)}\,e^{iqy}=
\frac{1}{\sqrt{4\lambda}} \int_{-\pi}^\pi dy \,e^{\frac{1}{4\lambda} (\cos y -1)}\,\cos(qy)\\
&\notag=\frac{1}{\sqrt{\lambda}} \int_{-\pi/2}^{\pi/2} dx\,e^{-\frac{1}{2\lambda} \sin^2 x} \,e^{2iqx}\\
& = \frac{\pi}{\sqrt{\lambda}} e^{-\frac{1}{4\lambda}} I_q \left(\frac{1}{4\lambda}\right)\,, 
\label{eq:O2terms}
\end{align}
with $q \in \mathbb{N}$, clearly $Z(\lambda)=\mathcal{O}_0(\lambda)$, while $\mathcal{O}_q(\lambda) = \mathcal{O}_{-q}(\lambda)$.

So how do we compute these observables in perturbation theory?
The way to proceed is first to complexify both the coupling constant, $\lambda = e^{i\theta} \vert \lambda \vert$, and the field variable, from $x\in [-\pi/2,\pi/2]$ to $z\in\mathbb{C}$, and the action, $S(z)$, becomes a meromorphic function of $z$. Secondly, we have to identify all the saddle points of the action in the whole complex plane $z\in\mathbb{C}$: 
\begin{equation}
S=\frac{1}{2\lambda} \sin^2 z\,,
\end{equation}
given by the usual Euler-Lagrange equation $dS / dz = 0$, whose solutions are simply
\begin{align}
 &z = 0 \qquad\rightarrow \qquad S\vert_{o.s.} =S_0= 0\,,\\
 &z=\pi/2\qquad \rightarrow \qquad S\vert_{o.s.}= S_1=
\frac{1}{2\lambda}\,.
\end{align}. 

The original integration contour $I = [-\pi/2, \pi/2]$ has real dimension $1$, so even after complexifing we still need to integrate over real dimension $1$ cycles, $\Sigma(\theta)$, that will depend on the argument $\theta$ of the coupling constant.
For each saddle point $z_i$, there is associated a unique integration cycle $\mathcal{J}_i$, called Lefschetz thimble or steepest descend path, defined by the flow equations:
\begin{align}
&\frac{\partial z}{\partial t} = - \overline{ \partial_z{S}(z)}\,,\label{eq:flow1}\\
&z(t\to-\infty) = z_i\,, \label{eq:flow2}
\end{align}
where $t$ is the ``time" along the thimble.
Thanks to the flow equations it is easy to show that the phase remains stationary along the thimbles
\begin{equation}
\mbox{Im} S(z) \vert_{\mathcal{J}_i} = \mbox{Im} S(z_i)\,.
\end{equation}
Equivalently the flows equations can be seen as hamiltonian equations for $(\mbox{Re} z, \mbox{Im} z)$ with $\mbox{Im} S$ as Hamiltonian, hence the stationarity of the phase\footnote{All these and all the following results are a direct consequence of the fact that $\mbox{Re} S$, being the real part of a holomorphic function, defines a perfect Morse function \cite{Witten:2010cx}.}.

The thimbles $\mathcal{J}_i$ are generically unbounded, even when the original integration contour is bounded. For this reason the convergence of the integral over a thimble is not guaranteed. We divide the complex $z$-plane into ``good" and ``bad" regions \cite{Witten:2010cx,Pham1983,Witten:2010zr}, good regions corresponds to $\mbox{Re} S(z) > 0$, while in bad regions $\mbox{Re} S(z)<0$.
The set of admissible Lefschetz thimbles, i.e. the ones whose asymptotic tails lay in the good regions, form a linearly independent and complete basis of integration cycles. 

As we dial the argument $\theta$ of the complexified coupling constant, every Lefschetz thimble $\mathcal{J}_i$ will deform smoothly and, for generic values of $\theta$, it will only pass through one saddle, i.e. its associated saddle $z_i$ from equation (\ref{eq:flow2}).
For specific values of $\theta$, precisely at Stokes lines, these contours will also pass through a subset of other saddles.
A generic integration cycle, $\Sigma(\theta)$, on which the integral converges, can be written as a sum over thimbles, so that $\Sigma(\theta)$ passes (in principle) from all the critical points:
\begin{equation}
\Sigma(\theta) = \sum_{i} n_i \mathcal{J}_i\,,\label{eq:decomp}
\end{equation}
where $n_i$ are integers.

\begin{figure}[tbp]
\centering
\includegraphics[width=0.9\textwidth]{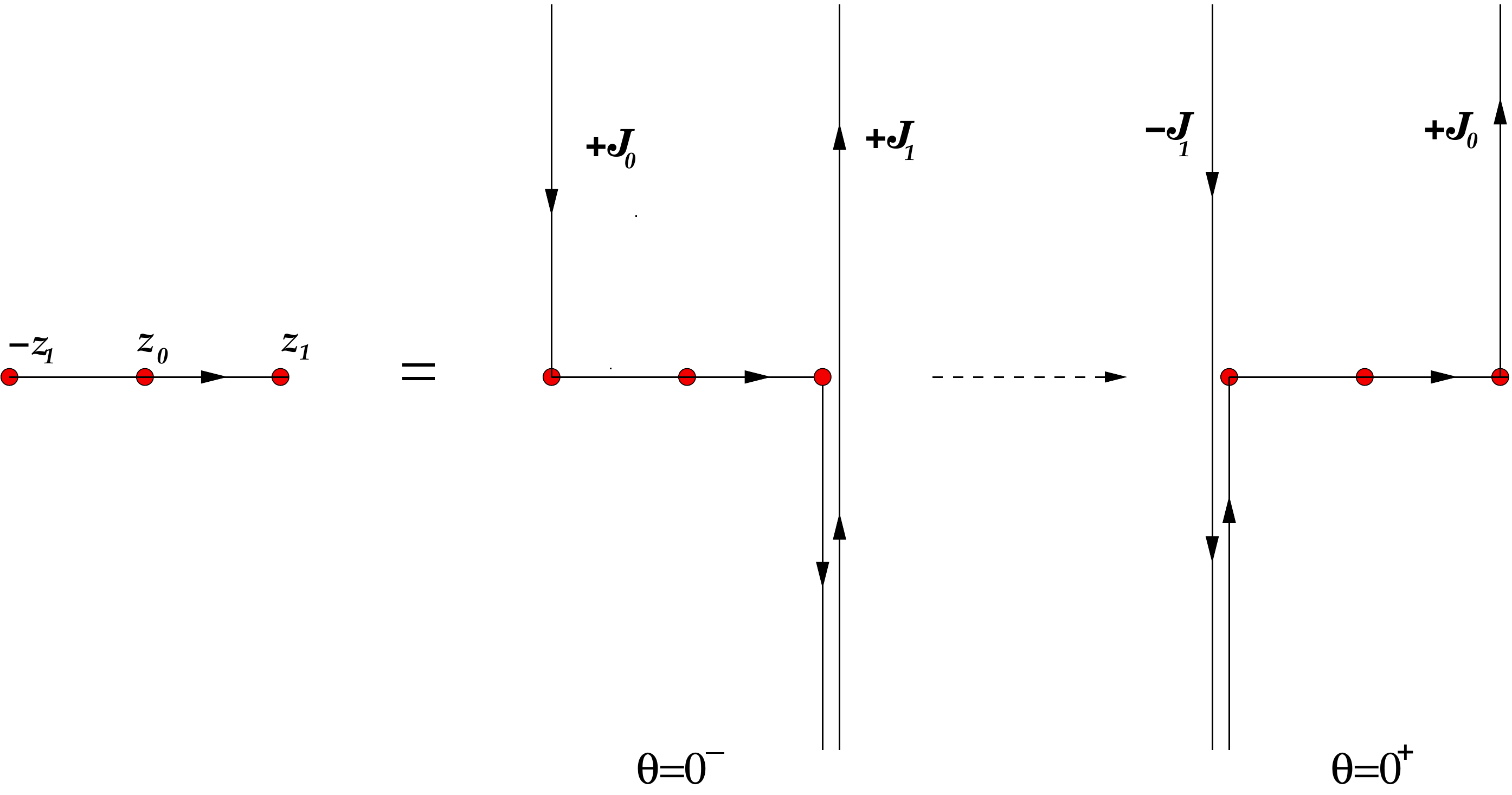}
\caption{The original integration cycle as a linear combination of Lefschetz thimbles at $\theta=0^-$ and   $\theta=0^+$. $\theta=0$ is a Stokes line. 
   }
 \label{fig:thimbles}
\end{figure}

These coefficients $n_i$ will jump precisely when $\theta$ crosses a Stokes line.
For example, in the case at hand, the original integration contour $I = [-\pi/2,\pi/2]$ can be written in two different ways, depending how we are approaching the real coupling case $\theta = 0$:
   \begin{align}
   \label{cycle-zero-one}
I = \left[- \frac{\pi}{2}, \frac{\pi}{2} \right]   \longrightarrow \Sigma = 
 \left\{ \begin{array}{l} 
   {\cal J}_0(0^{-})   + {\cal J}_1 (0^{-}) \cr
     {\cal J}_0 (0^{+} )- {\cal J}_1( 0^{+} ) 
\end{array} \right.
\end{align}
as shown in Figure \ref{fig:thimbles}.

If we were able to compute exactly the integrals along the different cycles, we would obtain an exact result for our observables.
Unfortunately, as usual, we cannot do that and we can only approximate each integral using perturbation theory around each different saddle.

The contribution to our observables (\ref{eq:O2terms}), coming from perturbation theory around the trivial vacuum $z=0$, is given by the asymptotic series
\begin{equation}
\mathfrak{Z}_0(\lambda,q) = e^{-S_0} \Phi_0 (\lambda,q)\,,
\end{equation}
and
\begin{equation}
\Phi_0(\lambda,q) = \sqrt{2\pi} \sum_{n=0}^\infty a^{(0)}_n(q) \,\lambda^n= \sqrt{2\pi} \sum_{n=0}^\infty \frac{(\frac{1}{2}+q)_n(\frac{1}{2}-q)_n}{n!} (2\lambda)^n\,,\label{eq:Phi0}
\end{equation}
where $(x)_n =  \Gamma(x+n)/ \Gamma(x)$ denotes the Pochhammer symbol.
As expected the perturbative series is diverging of Gevrey-1 type. Furthermore the $a_n$ are non-alternating in sign denoting the presence of a Stokes line for $\arg (\lambda)=0$.

Similarly, the perturbative contribution to (\ref{eq:O2terms}), coming from the non-perturbative saddle point $z=\pi$, is given by
\begin{equation}
\mathfrak{Z}_1(\lambda,q) = e^{-S_1} \Phi_1 (\lambda,q)\,,
\end{equation}
and
\begin{equation}
\Phi_1(\lambda,q) = \sqrt{2\pi} \sum_{n=0}^\infty a^{(1)}_n(q)  \,\lambda^n= \sqrt{2\pi} \sum_{n=0}^\infty \frac{(\frac{1}{2}+q)_n(\frac{1}{2}-q)_n}{n!} (-2\lambda)^n\,,
\end{equation}
which is once again of Gevrey-1 type but with coefficients $a^{(1)}_n$ alternating in sign, hallmark that $\arg (\lambda)=0$ is not a Stokes line for $\Phi_1(\lambda,q) $, while now  $\arg (\lambda)=\pi$ becomes the singular direction.

The semiclassical expansion for $\mathcal{O}_q$ can be represented with a two-terms trans-series
\begin{equation}
\mathcal{O}_q(\lambda,\sigma_0,\sigma_1) = \sigma_0 \mathfrak{Z}_0(\lambda,q) + \sigma_1 \mathfrak{Z}_1(\lambda,q) \,,
\label{eq:O2terms}
\end{equation}
where the $\sigma_i$ are the usual trans-series parameters.
We stress that a precise decomposition of the path integral will give us some definite values for the $\sigma_i$, whose role is exactly the same as the $n_i$ coefficients of the Lefschetz thimbles.
It is useful to keep (\ref{eq:O2terms}) with generic parameters $\sigma_i$ and study its resurgence properties and only at the end see the connection with the geometric structure of the path integral in terms of thimbles.

We can easily obtain the Borel transform of the above series
\begin{align}
\hat{\Phi}_0(\zeta,q) &= \sqrt{2\pi}\,_2F_1\left(q+\frac{1}{2},\frac{1}{2}-q,1\Big| 2\zeta\right) \,,\\
\hat{\Phi}_1(\zeta,q) &= \sqrt{2\pi}\,_2F_1\left(q+\frac{1}{2},\frac{1}{2}-q,1\Big| -2\zeta\right) \,,
\end{align} 
where $_2F_1$ is the hypergeometric function.
Both $\hat{\Phi}_0(\zeta,q)$ and $\hat{\Phi}_1(\zeta,q)$ define simple resurgent functions, with branch cuts respectively for $\zeta\in[1/2,+\infty)$ and $ \zeta\in (-\infty,-1/2]$.

Let us focus for the moment on the singular direction $\theta=0$, the discussion for $\theta = \pi$ will be exactly the same just by replacing $\Phi_0$ with $\Phi_1$.
We know the full discontinuity across the cut for the hypergeometric function \cite{NIST}:
\begin{eqnarray}
~_2F_1\left(a, b,c \Big|  \zeta+i\varepsilon\right)-~_2F_1\left(a,b,c \Big| \zeta-i\varepsilon\right)=  \frac{2 \pi i\,\Gamma(c)}{\Gamma(a) \Gamma(b)} ~_2F_1\left(c-a, c-b,1 \Big| 1-\zeta\right)\,, \qquad
\label{hyper-disc}
\end{eqnarray}
valid for $a+b=c$.
Using this relation, we can obtain the full discontinuity across the cut for $\hat{\Phi}_0$:
\begin{equation}
{\hat{\Phi}}_0(\zeta+i\epsilon,q)-{\hat{\Phi}}_0(\zeta-i\epsilon,q) = \frac{2 \pi i}{\Gamma(q+\frac{1}{2})\Gamma(\frac{1}{2}-q)} \sqrt{2 \pi} \,_2F_1\left(q+\frac{1}{2},\frac{1}{2}-q,1 \Big| 1-2\zeta\right)\,.
\end{equation}

\begin{figure}[tbp]
\centering
\includegraphics[width=1\textwidth]{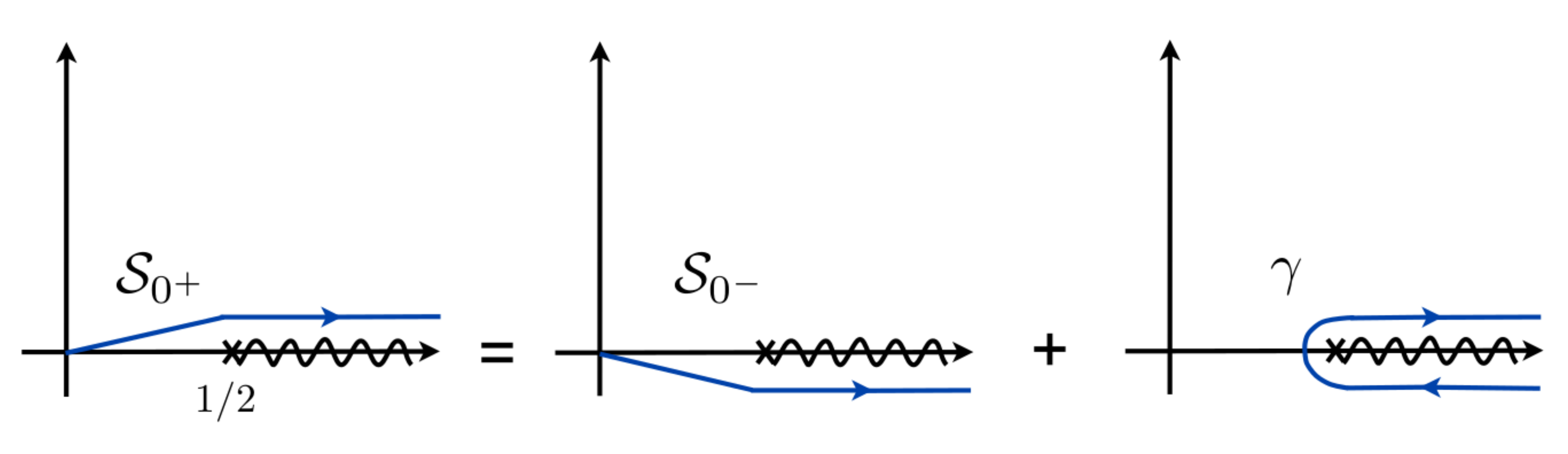}
\caption{The right Borel resummation can be rewritten as the sum of the left Borel resummation plus the contribution coming from the Hankel contour $\gamma$, coming from $t\to-\infty$, circling around the branch cut starting at $t=1/2$ and going back to $+\infty$.}
\label{fig:Hankel}
\end{figure}

Since we just obtained the full discontinuity for ${\hat{\Phi}}_0$ along the positive real axis, we can easily compute the Stokes automorphism and the full alien derivative algebra.
We proceed as illustrated in Fig.~\ref{fig:Hankel}: the difference of the right and left Borel resummation  can be written as an integral over the Hankel contour $\gamma$ which starts at $\infty$ below the imaginary axis, then circles  the singular point at $t=1/2$, and then goes back to $\infty$ above the imaginary axis:
\begin{align}
({\cal S}_{0^+}   - {\cal S}_{ 0^-} ) \Phi_{0} (\lambda,q) & = 
 \frac{\sqrt{2\pi}}{\lambda} \int_\gamma dt\,
  e^{- t/\lambda }   ~_2F_1\left(q+\frac{1}{2},\frac{1}{2}-q, 1 ;  2t \right)  \cr
& = \frac{\sqrt{2\pi}}{\lambda}   \int_{1/2}^\infty dt\,e^{-t/\lambda }\left[~_2F_1\left(q+\frac{1}{2},\frac{1}{2}-q, 1, 2t +i\varepsilon\right)-(i\epsilon\to- i\epsilon)\right] \cr
&= \frac{\sqrt{2\pi}}{\lambda}   \int_{1/2}^\infty dt \, e^{-  t/\lambda}\, \frac{2\pi i}{\Gamma(q+\frac{1}{2})\Gamma(\frac{1}{2}-q)}  ~_2F_1\left(q+\frac{1}{2},\frac{1}{2}-q, 1, 1-2t \right) \cr
&= \frac{2\pi i}{\Gamma(q+\frac{1}{2})\Gamma(\frac{1}{2}-q)} \frac{\sqrt{2\pi} e^{-1/(2 \lambda) }}{\lambda}   \int_{0}^\infty dt \, e^{-  t/\lambda}\,   ~_2F_1\left(q+\frac{1}{2},\frac{1}{2}-q, 1, -2t \right) \cr
&= \frac{2\pi i}{\Gamma(q+\frac{1}{2})\Gamma(\frac{1}{2}-q)}  e^{-1/(2 \lambda) }  {\cal S}_{0}  \Phi_{1} (\lambda,q) \,.
\label{disc3} 
\end{align}

We can combine our definition of the Stokes automorphism (\ref{eq:Stokes}), represented as difference of later Borel resummations, with the expression for the alien derivative as the logarithm of $\mathfrak{S}_\theta$ (\ref{eq:expAlien}), and
thanks to the equations (\ref{eq:1Sing}-\ref{eq:AlienSimple}), we obtain the only non-trivial alien derivatives along the Stokes line $\theta=0$
\begin{equation}
\Delta_{\frac{1}{2}} {\hat{\Phi}}_0 (\zeta,q) = 2i\sin(\pi (q+1/2)) {\hat{\Phi}}_1 (\zeta,q)\,,
\end{equation}
where we made use of the known formula $\Gamma(a)\Gamma(1-a) = \pi/ \sin(\pi a)$ to rewrite
the Stokes multiplier $A_{1/2}= 2\pi i/ (\Gamma(1/2+q)\Gamma(1/2-q))=2i\sin(\pi (q+1/2))$.
A similar story holds for $\theta = \pi$ where the only singular point is $\zeta=-1/2$, it is now $ {\hat{\Phi}}_1(\zeta,q)$ to have a non-trivial alien derivative
\begin{equation}
\Delta_{-\frac{1}{2}} {\hat{\Phi}}_1(\zeta,q) = 2i\sin(\pi (q+1/2)) {\hat{\Phi}}_0(\zeta,q)\,,
\end{equation}
and we note that the two Stokes constants are equal $A_{-1/2}=A_{1/2}$.

We can summarise the entire resurgent algebra for the problem at hand
\begin{align}
&\notag \Delta_{\frac{1}{2}} {\hat{\Phi}}_0(\zeta,q) = 2i (-1)^q {\hat{\Phi}}_1(\zeta,q) \,,\qquad\,\Delta_{-\frac{1}{2}}{\hat{\Phi}}_0(\zeta,q) = 0\,, \\
&\Delta_{\frac{1}{2}} {\hat{\Phi}}_1(\zeta,q)= 0\,,\qquad\qquad\qquad \qquad\Delta_{-\frac{1}{2}} \hat{\Phi}_1(\zeta,q) = 2i (-1)^q \, {\hat{\Phi}}_0(\zeta,q)\,,
\label{eq:ResAlg0d}
\end{align}
with all the other alien derivatives being zero.
This algebra is precisely of the form studied before in (\ref{eq:2TermResAlgebra}), the Stokes constants for these two-terms trans-series are $A_{1/2} = A_{-1/2}=2i(-1)^q$, when $q\in\mathbb{N}$.

Having the explicit expressions for the perturbative and non-perturbative saddle coefficients, we can check how the low orders coefficients of the NP physics are encoded in the large orders perturbative ones. Taking (\ref{eq:Phi0}) and changing variable to $z=1/\lambda$ we get
\begin{equation}
\Phi_0(z,q) = \sqrt{2\pi} +\sqrt{2\pi}\sum_{n=0}^\infty c^{(0)}_n(q) \,z^{-n-1}= \sqrt{2\pi} +\sqrt{2\pi}\sum_{n=0}^\infty \frac{(\frac{1}{2}+q)_{n+1}(\frac{1}{2}-q)_{n+1}}{(n+1)!} (2/z)^{n+1}\,.
\end{equation}
We can easily extract the high orders behaviour of the perturbative coefficients $c^{(0)}_n$ and write it in the suggestive form
\begin{align}
c^{(0)}_n\sim \frac{1}{2\pi i}\, &\notag\frac{2\pi i}{ \Gamma(1/2+q) \Gamma(1/2-q)}\, \frac{n!}{(1/2)^{n+1}}\,\times \\
&\times \left(
1+(4q^2-1)\frac{1/2}{n}+\frac{16 q^4-40 q^2+9}{2}\frac{(1/2)^2}{n(n-1)}+...
\right)\,.
\end{align}
Comparing this expansion for $c^{(0)}_n$ with (\ref{eq:HigherOrderP}-\ref{eq:HigherOrderNP}) we can read precisely the lower orders coefficients $c^{(1)}_0,c^{(1)}_1,...$ of the perturbative expansion for the non-trivial saddle. Furthermore we can recover the instanton action $M=1/2$ and the Stokes constant $A_{1/2}=2\pi i/ (\Gamma(1/2+q) \Gamma(1/2-q))$ just obtained via alien calculus.
The perturbative vacuum knows everything about the non-perturbative saddle point and vice versa!

The action of the Stokes automorphisms $\mathfrak{S}_{0},\mathfrak{S}_{\pi}$ has been computed in (\ref{eq:StokesZero2terms}-\ref{eq:StokesPi2terms}) so from (\ref{eq:Stokes}) we can relate the left and right resummation of the full trans-series 
\begin{align}
\mathcal{S}_{0^+} \mathcal{O}_q(\lambda,\sigma_0,\sigma_1) &\label{eq:StokesOplus}= \mathcal{S}_{0^-} \mathcal{O}_q(\lambda,\sigma_0,\sigma_1+2i (-1)^q \sigma_0)\,,\\
\mathcal{S}_{\pi^+} \mathcal{O}_q(\lambda,\sigma_0,\sigma_1) &= \mathcal{S}_{\pi^-} \mathcal{O}_q(\lambda,\sigma_0+2i (-1)^q  \sigma_1,\sigma_1)\,.
\end{align}
Had we had started with the perturbative series $\mathfrak{Z}_0$ alone, by crossing the positive real axis (one of the two Stokes line), we would have automatically generated the recessive (exponentially suppressed) term $\mathfrak{Z}_1$, coming from perturbation theory around the non-trivial saddle-point.

We have finally come to the question: How do we assign an unambiguous sum to the trans-series (\ref{eq:O2terms}), corresponding to (\ref{eq:Z2terms}) and (\ref{eq:O2terms})?
If we compute a \textit{real} physical quantity in terms of a trans-series
\begin{equation}
\mathcal{O}(z)=\sum_j c_j \mathfrak{g}_j(z)\,,
\end{equation}
the reality of this observable translates into a reality condition for both the coefficients $c_j\in\mathbb{R}$ and the trans-monomials $\mathfrak{g}_j$:
\begin{equation}
\left(\mathcal{C}\,\mathcal{O}\right) (z) = \mathcal{O}(\bar{z})\,,
\end{equation}
where  $\mathcal{C}$ denotes the complex conjugation operator.

For the $0d$ model observables (\ref{eq:O2terms}) it is manifest that both the partition function $Z(\lambda)$ and the conformal primaries $\mathcal{O}_q(\lambda)$ should be real functions when $\lambda$ is real. 
We have already seen that if we resum the trans-series along a Stokes line, using either $\mathcal{S}_{0^+}$ or $\mathcal{S}_{0^-}$,  we will not obtain something real (\ref{eq:StokesOplus}), focusing for example on the positive real axis:
\begin{align}
&\mathcal{C} \left( \mathcal{S}_{0^+} \mathcal{O} \right) (z) \neq \left(\mathcal{S}_{0^+} \mathcal{O}\right)(\bar{z})\,,\\
&\mathcal{C} \left( \mathcal{S}_{0^-} \mathcal{O} \right) (z) \neq \left(\mathcal{S}_{0^-} \mathcal{O}\right)(\bar{z})\,.
\end{align}

This follows from the fact that on a singular direction the Stokes automorphism acts non-trivial on our trans-series, i.e. $\mathcal{S}_{0^+}\mathcal{O}\neq\mathcal{S}_{0^-}\mathcal{O}$, while instead the complex conjugation swaps the two lateral summations:
\begin{equation}
\mathcal{C} \circ \mathcal{S}_{0^+} = \mathcal{S}_{0^-} \circ \mathcal{C}\,,
\label{eq:CCS}
\end{equation}
which translates into
\begin{equation}
\mathcal{C} \circ \mathfrak{S}_0^{-1} =\mathfrak{S}_0 \circ \mathcal{C} \,.
\end{equation}
This means that the conjugation operator has to anti-commute with the alien derivative
\begin{equation}
\mathcal{C} \circ \dot{\Delta}_0 = - \dot{\Delta}_0 \circ \mathcal{C}\,,
\end{equation}
where $\dot{\Delta}_0=\sum_{\omega\in\Gamma_0} \dot{\Delta}_{\omega}$.
Across a Stokes line neither of the two lateral summations can possibly give a real resummation of our real observable.

To obtain a real resummation procedure we have to introduce what it is called median summation.
Firstly, since we express the alien derivative as the logarithm of the Stokes automorphism (\ref{eq:expAlien}), we can define non-integers power of $\mathfrak{S}$ simply by
\begin{equation}
\mathfrak{S}_\theta^\nu \doteq \exp  \left(\nu \sum_{\omega\in\Gamma_\theta} e^{-\omega\,z}\Delta_{\omega}\right)=\exp \left(\nu\,\dot{\Delta}_\theta\right)\,,
\end{equation} 
with $\nu \in \mathbb{C}$ while $\Gamma_\theta$ denotes the set of singular points along the $\theta$ direction.
Thanks to the our discussion above, we know now how complex conjugation acts on the non-integers power of the Stokes automorphism
\begin{equation}
\mathcal{C} \circ \mathfrak{S}_0^\nu = \mathfrak{S}_0^{-\nu} \circ \mathcal{C}\,.
\label{eq:CCFracStokes}
\end{equation}

\begin{Def}
We define the \textit{median resummation}
\begin{equation}
\mathcal{S}_{med} = \mathcal{S}_{0^-} \circ \mathfrak{S}_0^{-1/2} = \mathcal{S}_{0^+} \circ \mathfrak{S}_0^{1/2}\,,
\end{equation}
which, in contrast to $\mathcal{S}_{0^\pm}$, resum power series with real coefficients into real analytic functions of $z$, for $z\in\mathbb{R}^+$.
\end{Def}

The median resummation does precisely what we were expecting from our summation procedure
\begin{equation}
\mathcal{C} \left( \mathcal{S}_{med} \mathcal{O} \right) (z) = \left(\mathcal{S}_{med} \mathcal{O}\right)(\bar{z})\,,
\end{equation}
where we used equation (\ref{eq:CCFracStokes}) for $\nu=1/2$.

Let's go back to our explicit $0d$ example.
In this case it is pretty easy to compute $\mathfrak{S}_0^\nu$ since we know the full resurgent algebra (\ref{eq:ResAlg0d})
\begin{align}
\mathfrak{S}_0^{\nu} \mathcal{O}_q(\lambda,\sigma_0,\sigma_1) = \mathcal{O}_q(\lambda,\sigma_0,\sigma_1+2i(-1)^q \nu\,\sigma_0)\,.
\end{align}
We can finally notice that the median resummation for our observables leads to
\begin{align}
\left(\mathcal{O}_q\right)_{\mathbb{R}}(\lambda) &\notag= \mathcal{S}_{med} \mathcal{O}_q(\lambda,1,0)\\
&= \mathcal{S}_{0^+}  \mathcal{O}_q(\lambda,1,-i(-1)^q)= \mathcal{S}_{0^-}
 \mathcal{O}_q(\lambda,1,+i(-1)^q)\,,
 \end{align}
 which is exactly equivalent to the correct decomposition of the path integral in terms of Lefschetz thimbles, with the precise intersection numbers $n_i$, computed by Morse theory, necessary to write our original integration contour as a sum of steepest descent paths (\ref{eq:decomp}).
 
As we have seen before the original path integral does receive contribution from both thimbles, let's consider for example the contribution to the partition function coming from the perturbative vacuum thimble. Changing variables from the field variable $z$ to the action variable $u=S(z)$:
   \begin{align}
 \frac{1}{\sqrt \lambda} \int_{{\cal J}_0 (0^\mp) } e^{-\frac{1}{2 \lambda} \sin^2(z)}  
  &= \frac{2}{\sqrt \lambda} \int_0^{1/2} du \frac{e^{-u/\lambda}}{\sqrt {2 u (1-2u)}}  
 \mp \frac{2i}{\sqrt \lambda} \int_{1/2}^{\infty}  du \frac{e^{-u/\lambda}}{\sqrt {2 u (2u-1)}} \cr
&= \frac{2}{\sqrt \lambda} \int_0^{1/2} du \frac{e^{-u/\lambda}}{\sqrt {2 u (1-2u)}}  
 \mp i e^{- \frac{1}{2 \lambda} } \frac{2}{\sqrt \lambda} \int_{0}^{\infty} du \frac{e^{-u/\lambda}}{\sqrt {(2 u+1) 2u}}   \cr 
 &=  {\rm Re} {\cal S}_{0}    {\Phi}_0   \mp i e^{- \frac{1}{2 \lambda} } {\cal S}_{0}    {\Phi}_1 \label{eq:J0pm}
 \end{align}
where  ${\rm Re} {\cal S}_{0}    {\Phi}_0 $ is unambiguous.
The integral  that we identify with 
${\rm Re} {\cal S}_{0}    {\Phi}_0 $ is dominated by $ u \lesssim \lambda$ in the small $\lambda$ regime.  The procedure to obtain the perturbative expansion  ${\Phi}_0$ from this expression involves two steps.  First, we should extend the integration domain to $[0, \infty)$. Secondly, we Taylor expand 
$\frac{1}{\sqrt {(1-2u)}}  $ around the origin, and, performing the integration term by term, we will obtain the divergent asymptotic expansion  ${\Phi}_0$. The reason for the divergence is the use of the Taylor expansion beyond its radius of convergence\footnote{One can also obtain the (convergent)  strong coupling expansion from the integral representation in the  $\lambda \gg 1$ regime, by expanding the exponential into a power series in $\frac{1}{\lambda}$ and performing order by order integration.}.

We recover in different form the presence of a Stokes phenomenon for the perturbative series at $\theta=0$ together with a non-trivial Stokes automorphism (\ref{disc3}). The contributions to the partition function from the perturbative Lefschetz thimbles $\mathcal{J}_0(0^\pm)$ give rise to an ambiguity for real coupling, but of course this is not the only term to consider to get the full answer.
We know that the original domain of integration, $I=[-\pi/2,\pi/2]$, has to be written as a linear combination of $\mathcal{J}_0$ and $\mathcal{J}_1$ according to (\ref{cycle-zero-one}). It is only after adding the term arising from the $\mathcal{J}_1$ thimble, with precisely the right intersection number $n_1$ (function of $\theta$), that
we get an exact cancellation between the ambiguity coming from the resummation of the perturbation theory and the jump of the $\mathcal{J}_1$ contribution. Rather than coming from our resummation procedure,the jump of the $\mathcal{J}_1$ contribution is geometric in nature instead, and it is due to the different integration contours decompositions, as we approach $\theta=0^\pm$!

To conclude, the ambiguity in the imaginary part of the integration  $\int_{{\cal J}_0 (0^\pm) }$ is cancelled exactly by the  ambiguity in the prefactor of the $  \int_{{\cal J}_1 (0^\pm) } $ integral. 
The path integral, decomposed into thimbles, gives precisely the median resummation prescription from first principles, as a geometric construction, the ambiguity coming from the resummation of the perturbative expansion is intertwined with the 
jumps in the decomposition of the original contour of integration into steepest descent path, in the spirit of Morse theory.
  In this simple $0$d case, the median resummation is fairly easy to obtain and consists simply in taking the real part of the $\mathcal{S}_0$ resummation of the perturbation series around the vacuum alone (\ref{eq:J0pm}). The generic case when logarithm and branch cuts are present is much more complicated and has been discussed in meticulous details in \cite{Aniceto:2013fka}.
  
 \section{Outlook} 
  \label{sec:Out}
  
From the analysis carried out in the previous Section, we can draw a simple analogy between finite dimensional integrals, where the relation between saddle points approximation and trans-series expansion is well established \cite{berry1990hyperasymptotics,berry1991hyperasymptotics}, and the path integral formulation of QM and QFT, to understand why physical observables should be obtained as trans-series expansions of simple resurgent functions. In a semiclassical path integral calculation, we should first look for all the finite action, classical solutions to the equation of motion, in a suitable complexification of our fields space, then, by deforming the original ``contour" of integration, we should add up all these exponentially suppressed contributions, together with the remaining fluctuations on top of them, i.e. zero-, quasi-zero- and gaussian-modes.
   
We have to stress that, while in euclidean QM and QFT, it is natural to expect instantons contributions in an euclidean path integral calculation (simply because we can construct these finite action non-perturbative saddles), in real time QM and QFT, on the other hand, it is not. The ``weight" of a classical configuration, in the real time path integral, is given by $e^{i S}$, so it is not clear how a saddle points expansion could give rise to exponentially suppressed terms, i.e. the energy splitting in a double-well  QM. As it turns out \cite{Tanizaki:2014xba,Cherman:2014sba}, even if we complexify all our fields and look for more generic, real time saddle points, we still do not find instanton-like configurations. To find smooth and exponentially suppressed instantons, it seems that we have to complexify both the fields and the time variable! Note that there is no need to go all the way to imaginary time $t\to -i \tau$, i.e. the euclidean formulation, as soon as we work a little bit off the real time $t\to t \,e^{i \epsilon}$ (very much likely the $+i \epsilon$-prescription for the propagators), regular, finite and real action instantons appear as solutions to the complexified classical equations of motion.

Going back to the Euclidean case, in QM \cite{Bogomolny:1980ur,ZinnJustin:1981dx,ZinnJustin:1982td} and QFT \cite{Cherman:2014ofa}, instanton-anti-instanton type of amplitudes possess unambiguous real parts and two-fold ambiguous imaginary parts, necessary to ``cure" the ambiguity coming from the resummation of perturbation theory around the perturbative vacuum alone.
 The cancellation of the imaginary parts in path integral examples is essentially the same as for ordinary integrals.  However, in QFT there are also real unambiguous contributions to observables from NP-saddles.

Of course, in semiclassically calculable regimes of QFTs and in QM,  there are infinitely many saddle points and, even if a mathematically rigorous definition for the path integral measure is lacking, we can still ask the question: is it possible to write our original, infinite dimensional, ``contour" of integration, i.e. the space of fields, as an infinite sum of nicer, infinite dimensional thimbles, living in a suitable complexification of the original space of fields? 

We can try to use the (complexified) action as a Morse function on the complexification of the space of fields, and, as a generalisation of the flow equations (\ref{eq:flow1})-(\ref{eq:flow2}) described above, we get that a thimble is a solution to
\begin{align*}
&\frac{\partial \phi(x,t) }{\partial t} = -\overline{ \frac{ \delta{S[\phi]}}{\delta \phi(x,t)}}\,,\\
& \lim_{t\to-\infty} \phi(x, t) = \phi_{cl}(x)\,,\\
&\left.\frac{\delta S[\phi]}{\delta \phi (x,t)}\right\vert_{\phi(x,t)=\phi_{cl}(x)}=0   \,.
\end{align*}
 
The flow\footnote{Generically these flow equations are parabolic PDEs so the flow is actually only a semi-flow.} originates, at $t\to-\infty$, from a classical solution, $\phi_{cl}(x)$, to the complexified equation of motion and, as we increase a little the flow time $t$, we start moving in the space of fields accordingly to the eigenvectors and eigenvalues of the quadratic fluctuations operator $\delta^2 S/ \delta \phi^2 \vert_{\phi_{cl}}$.
The thimbles are infinite dimensional algebraic varieties and when the theory is regularised on a finite lattice with a finite size, they become finite dimensional algebraic varieties \cite{Cristoforetti:2012su}.

In few lucky cases, namely 3d Chern-Simons \cite{Witten:2010cx} and QM in phase space \cite{Witten:2010zr},
these flow equations become elliptic PDEs and can be solved. The original ``contour" of integration in the path integral can be then rewritten in terms of thimbles living in a complexification of the original fields space, and each thimble is associated to one (away from Stokes lines) classical solution to the complexified equations of motion in the same spirit of Section \ref{sec:Median}.

For generic QFTs we do not know if the original ``contour" of integration of the path integral can be decomposed as a sum of thimbles, but both Floer homology for parabolic flows \cite{Salamon} and our trans-series expansion for physical observables seems to suggest that this is indeed the case.
The infinite dimensional analog of Lefschetz thimbles in QFT and QM is far from being completely understood \cite{Harlow:2011ny}, but it looks exactly the right setup to understand why physical observables can be obtained as trans-series expansions for simple resurgent functions.

\section{Acknowledgment}
I am very grateful to Mithat \"Unsal for introducing me to the fascinating subject of resurgence and for countless discussions, and to Gerald Dunne for his invaluable help, careful proof reading of these notes and for being a continuous source of new ideas. 
I would also like to thank Ines Aniceto, Gokce Basar, Aleksey Cherman, Hugh Osborn, Slava Rychkov, David Sauzin and Kenny Wong for stimulating and enjoyable discussions. I am grateful for the support of European Research Council Advanced Grant No. 247252, Properties and Applications of the Gauge/Gravity Correspondence.



\begin{thebibliography}{10}

\bibitem{Dyson:1952tj}
F.~Dyson, {\it {Divergence of perturbation theory in quantum electrodynamics}},
  Phys.Rev. {\bf 85} (1952) 631--632.

\bibitem{Hurst:1952zh}
  C.~A.~Hurst,
  {\it{The Enumeration of Graphs in the Feynman-Dyson Technique}},
  Proc.\ Roy.\ Soc.\ Lond.\ A {\bf 214} (1952) 44.

\bibitem{Bender:1976ni}
  C.~M.~Bender and T.~T.~Wu,
  { \it{Statistical Analysis of Feynman Diagrams}},
  Phys.\ Rev.\ Lett.\  {\bf 37} (1976) 117.


\bibitem{Lipatov:1976ny}
L.~Lipatov, {\it {Divergence of the Perturbation Theory Series and the
  Quasiclassical Theory}},  Sov.Phys.JETP {\bf 45} (1977) 216--223.
  
  
\bibitem{Bender:1969si}
C.~M. Bender and T.~T. Wu, {\it {Anharmonic Oscillator}},  Phys. Rev. {\bf 184}
  (1969) 1231.
  
\bibitem{ZinnJustin:1980uk}
J.~Zinn-Justin, {\it {Perturbation Series at Large Orders in Quantum Mechanics
  and Field Theories: Application to the Problem of Resummation}},  Phys. Rept.
  {\bf 70} (1981) 109.


\bibitem{Bender:1990pd}
C.~M. Bender and T.~Wu, {\it {Anharmonic Oscillator 2: A Study of Perturbation
  Theory in Large Order}},  Phys. Rev. {\bf D7} (1973) 1620.
  

\bibitem{Marino:2007te}
M.~Mari{\~n}o, R.~Schiappa and M.~Weiss, {\it {Nonperturbative Effects and the
  Large--Order Behavior of Matrix Models and Topological Strings}},  Commun.
  Num. Theor. Phys. {\bf 2} (2008) 349.

\bibitem{Pasquetti:2009jg}
S.~Pasquetti and R.~Schiappa, {\it {Borel and Stokes Nonperturbative Phenomena
  in Topological String Theory and $c=1$ Matrix Models}},  Annales Henri
  Poincar{\'e} {\bf 11} (2010) 351.

\bibitem{Marino:2006hs}
M.~Mari{\~n}o, {\it {Open String Amplitudes and Large--Order Behavior in
  Topological String Theory}},  JHEP {\bf 0803} (2008) 060.



\bibitem{Dingle_asymptotics}
R.~B. Dingle, {\em Asymptotic Expansions: Their Derivation and Interpretation}.
\newblock Academic Press, 1973.

  
\bibitem{Grassi:2014cla}
  A.~Grassi, M.~Marino and S.~Zakany,
  {\it{Resumming the string perturbation series}},
  arXiv:1405.4214 [hep-th].

\bibitem{Heller:2013fn}
  M.~P.~Heller, R.~A.~Janik and P.~Witaszczyk,
  {\it{Hydrodynamic Gradient Expansion in Gauge Theory Plasmas}},
  Phys.\ Rev.\ Lett.\  {\bf 110} (2013) 21,  211602.

  \bibitem{Ecalle:1981}
J.~Ecalle, {\em Les Fonctions Resurgentes}, vol.~I - III.
\newblock Publ. Math. Orsay, 1981.

\bibitem{Voros:1983}
A.~Voros, {\it {The Return of the Quartic Oscillator: The Complex WKB Method}},
   Ann. Inst. Henri Poincar{\'e} {\bf 39} (1983) 211.
   
   
\bibitem{Voros:1993an}
A.~Voros, {\it {R{\'e}surgence Quantique}},  Ann. Inst. Fourier {\bf 43} (1993)
  1509.
  
\bibitem{Delabaere:1999ef}
E.~Delabaere and F.~Pham, {\it {Resurgent Methods in Semi--Classical
  Asymptotics}},  Ann. Inst. Henri Poincar{\'e} {\bf 71} (1999) 1.
  
  
 \bibitem{Jentschura:2004jg}
U.~D. Jentschura and J.~Zinn-Justin, {\it {Instantons in Quantum Mechanics and
  Resurgent Expansions}},  Phys. Lett. {\bf B596} (2004) 138.
 

  \bibitem{Marino:2012zq}
M.~Mari{\~n}o, {\it {Lectures on Non--Perturbative Effects in Large $N$ Gauge
  Theories, Matrix Models and Strings}},
 Fortsch.\ Phys.\  {\bf 62} (2014) 455.


\bibitem{Aniceto:2014hoa}
  I.~Aniceto, J.~G.~Russo and R.~Schiappa,
  {\it{Resurgent Analysis of Localizable Observables in Supersymmetric Gauge Theories}},
  arXiv:1410.5834 [hep-th].



\bibitem{Aniceto:2011nu}
I.~Aniceto, R.~Schiappa and M.~Vonk, {\it {The Resurgence of Instantons in
  String Theory}},  Commun. Num. Theor. Phys. {\bf 6} (2012) 339.


\bibitem{Santamaria:2013rua}
  R.~C.~Santamaria, J.~D.~Edelstein, R.~Schiappa and M.~Vonk,
  {\it{Resurgent Transseries and the Holomorphic Anomaly}},
  arXiv:1308.1695 [hep-th].
  

\bibitem{Couso-Santamaria:2014iia}
  R.~C.~Santamaria, J.~D.~Edelstein, R.~Schiappa and M.~Vonk,
  {\it{Resurgent Transseries and the Holomorphic Anomaly: Nonperturbative Closed Strings in Local CP2}},
  arXiv:1407.4821 [hep-th].

  
  \bibitem{Dunne:2012ae}
G.~V. Dunne and M.~{\"U}nsal, {\it {Resurgence and Trans--Series in Quantum
  Field Theory: The ${\mathbb{CP}}^{N-1}$ Model}},  JHEP {\bf 1211} (2012) 170.

\bibitem{Dunne:2012zk}
G.~V. Dunne and M.~{\"U}nsal, {\it {Continuity and Resurgence: Towards a
  Continuum Definition of the ${\mathbb{CP}}^{N-1}$ Model}},  Phys. Rev. {\bf D87}
  (2013) 025015.

\bibitem{Argyres:2012vv}
P.~Argyres and M.~{\"U}nsal, {\it {A Semiclassical Realization of Infrared
  Renormalons}},  Phys. Rev. Lett. {\bf 109} (2012) 121601.

\bibitem{Argyres:2012ka}
P.~C. Argyres and M.~{\"U}nsal, {\it {The Semi--Classical Expansion and
  Resurgence in Gauge Theories: New Perturbative, Instanton, Bion, and
  Renormalon Effects}},  JHEP {\bf 1208} (2012) 063.

\bibitem{'tHooft:1977am}
G.~'t~Hooft, {\it {Can We Make Sense Out of Quantum Chromodynamics?}},
  Subnucl. Ser. {\bf 15} (1979) 943.

\bibitem{Beneke:1998ui}
M.~Beneke, {\it {Renormalons}},  Phys. Rept. {\bf 317} (1999) 1.


\bibitem{Sauzin1}
D.~{Sauzin}, {\it {Resurgent functions and splitting problems}},
  arXiv:0706.0137 [math.DS].
  
  \bibitem{Sauzin2}
D.~{Sauzin}, {\it {Introduction to 1-summability and resurgence }},
  arXiv:1405.0356 [math.DS].


\bibitem{Cherman:2013yfa}
A.~Cherman, D.~Dorigoni, G.~V. Dunne and M.~{\"U}nsal, {\it {Resurgence in QFT:
  Unitons, Fractons and Renormalons in the Principal Chiral Model}},
  Phys.\ Rev.\ Lett.\  {\bf 112} (2014) 021601.


\bibitem{Edgar:2008ga}
G.~A. Edgar, {\it {Transseries for Beginners}},  Real Anal. Exchange {\bf 35}
  (2009) 253.


\bibitem{Dunne:2013ada}
G.~V. Dunne and M.~{\"U}nsal, {\it {Generating Energy Eigenvalue Trans--Series
  from Perturbation Theory}},
   Phys.\ Rev.\ D {\bf 89} (2014) 041701.


\bibitem{Dunne:2014bca}
  G.~V.~Dunne and M.~Unsal,
  {\it {Uniform WKB, Multi-instantons, and Resurgent Trans-Series,}}
  arXiv:1401.5202 [hep-th].


\bibitem{Bogomolny:1980ur}
E.~Bogomolny, {\it {Calculation of Instanton---Anti--Instanton Contributions in
  Quantum Mechanics}},  Phys. Lett. {\bf B91} (1980) 431.

\bibitem{ZinnJustin:1981dx}
J.~Zinn-Justin, {\it {Multi--Instanton Contributions in Quantum Mechanics}},
  Nucl. Phys. {\bf B192} (1981) 125.

\bibitem{ZinnJustin:1982td}
J.~Zinn-Justin, {\it {Multi--Instanton Contributions in Quantum Mechanics 2}},
  Nucl. Phys. {\bf B218} (1983) 333.


\bibitem{Cherman:2014ofa}
  A.~Cherman, D.~Dorigoni and M.~Unsal,
  {\it{Decoding perturbation theory using resurgence: Stokes phenomena, new saddle points and Lefschetz thimbles,}}
  arXiv:1403.1277 [hep-th].


	 
\bibitem{Delabaere:2006ed}
E.~Delabaere, {\it {Effective Resummation Methods for an Implicit Resurgent
  Function}},
 arXiv:0602026 [math-ph].


\bibitem{costin2008asymptotics}
O.~Costin, {\em Asymptotics and Borel summability}.
\newblock CRC Press, 2008.

\bibitem{Gair:2011mr}
  J.~Gair, N.~Yunes and C.~M. Bender,
  {\it {Resonances in Extreme Mass-Ratio Inspirals: Asymptotic and Hyperasymptotic Analysis,}}
  J.\ Math.\ Phys.\  {\bf 53} (2012) 032503.
  
  
\bibitem{Garoufalidis:2010ya}
S.~Garoufalidis, A.~Its, A.~Kapaev and M.~Mari{\~n}o, {\it {Asymptotics of the
  Instantons of Painlev{\'e} I}},  Int. Math. Res. Notices {\bf 2012} (2012)
  561.


\bibitem{Schiappa:2013opa}
R.~Schiappa and R.~Vaz, {\it {The Resurgence of Instantons: Multi--Cut Stokes
  Phases and the Painlev{\'e} II Equation}},
  \  Commun.\ Math.\ Phys.\  {\bf 330} (2014) 655.


\bibitem{Basar:2013eka}
  G.~Basar, G.~V.~Dunne and M.~Unsal,
  {\it{Resurgence theory, ghost-instantons, and analytic continuation of path integrals,}}
  JHEP {\bf 1310} (2013) 041.
  
  
\bibitem{Aniceto:2013fka}
  I.~Aniceto and R.~Schiappa,
  {\it {Nonperturbative Ambiguities and the Reality of Resurgent Transseries,}}
  arXiv:1308.1115 [hep-th].
  
\bibitem{berry1993unfolding}
	M.~V.~Berry and C.~J.~Howls, {\it{Unfolding the high orders of asymptotic expansions with coalescing saddles: singularity theory, crossover and duality}},
	 Proc. R. Soc. A {\bf{443}} (1993), 107-126.
  

\bibitem{Cherman:2014xia}
  A.~Cherman, P.~Koroteev and M.~†nsal,
  {\it{Resurgence and Holomorphy: From Weak to Strong Coupling}},
  arXiv:1410.0388 [hep-th].

  
\bibitem{Witten:2010cx}
E.~Witten, {\it {Analytic Continuation Of Chern-Simons Theory}},
 arXiv:1001.2933 [hep-th].

\bibitem{Pham1983}
F.~Pham, {\it Vanishing homologies and the n variable saddlepoint method},
 Proc. Symp. Pure Math. {\bf 2} (1983), no.~40 319--333.


  \bibitem{Witten:2010zr}
E.~Witten, {\it {A New Look At The Path Integral Of Quantum Mechanics}},
  arXiv:1009.6032 [hep-th].
  
\bibitem{NIST}
F.~W. Olver, D.~W. Lozier, R.~F. Boisvert, and C.~W. Clark, {\em {NIST Handbook
  of Mathematical Functions}}.
\newblock Cambridge University Press, 2010.


\bibitem{berry1990hyperasymptotics}
	M.~V.~Berry and C.~J.~Howls, {\it{Hyperasymptotics}},
	 Proc. R. Soc. A {\bf{430}} (1990), 653-668.


\bibitem{berry1991hyperasymptotics}
	M.~V.~Berry and C.~J.~Howls, {\it{Hyperasymptotics for integrals with saddles}},
	 Proc. R. Soc. A {\bf{434}} (1991), 657-675.

\bibitem{Tanizaki:2014xba}
  Y.~Tanizaki and T.~Koike,
  {\it{Real-time Feynman path integral with Picard--Lefschetz theory and its applications to quantum tunneling}},
  Annals Phys.  {\bf 351} (2014) 250.
  

\bibitem{Cherman:2014sba}
  A.~Cherman and M.~Unsal,
 {\it{Real-Time Feynman Path Integral Realization of Instantons}},
  arXiv:1408.0012 [hep-th].


\bibitem{Cristoforetti:2012su}
  M.~Cristoforetti,  F.~Di Renzo and L.~Scorzato,
 { \it{New approach to the sign problem in quantum field theories: High density QCD on a Lefschetz thimble}},
  Phys.\ Rev.\ D {\bf 86} (2012) 074506
 
   \bibitem{Salamon}
D.~A.~Salamon and J.~Weber, {\it {Floer homology and the heat flow }},
  arXiv:math/0304383 [math.SG].
 
\bibitem{Harlow:2011ny}
D.~Harlow, J.~Maltz, and E.~Witten, {\it {Analytic Continuation of Liouville
  Theory}},  JHEP {\bf 1112} (2011) 071.

\end{thebibliography}
\end{document}